\documentclass[usenatbib]{mn2e}
\usepackage{amssymb,graphicx,natbibmnfix,times}

\newcommand{\Msun}{\mbox{$M_\odot$}}
\newcommand{\xHeIII}{$x_{\rm{HeIII}}$}
\newcommand{\xHeII}{$x_{\rm{HeII}}$}
\newcommand{\Lya}{Ly$\alpha$ }
\newcommand{\lya}{Ly$\alpha$}

\begin{document}

\title[He II reionization and radiation background]{Semi-numeric simulations of helium reionization and the fluctuating radiation background}

\author[K.~L.~Dixon, et al.]{Keri L. Dixon$^{1,2}$\thanks{email: K.Dixon@sussex.ac.uk}, Steven R. Furlanetto$^{2}$ \& Andrei Mesinger$^{3}$
\\
$^1$Astronomy Centre, University of Sussex, Falmer, Brighton BN1 9QH, United Kingdom \\
$^2$Department of Physics and Astronomy, University of California, Los Angeles, CA 90095, USA \\
$^3$Scuola Normale Superiore, Piazza dei Cavalieri 7, 56126 Pisa, Italy
}

\date{\today} \pubyear{2013} \volume{000}
\pagerange{1} \twocolumn 

\voffset-.6in
\maketitle

\begin{abstract}  

Recent He~II Lyman-$\alpha$ forest observations from $2.0 \la z \la 3.2$ show large fluctuations in the optical depth at $z \gtrsim 2.7$. These results point to a fluctuating He-ionizing background, which may be due to the end of helium reionization of this era. We present a fast, semi-numeric procedure to approximate detailed cosmological simulations. We compute the distribution of dark matter halos, ionization state of helium, and density field at $z = 3$ in broad agreement with recent simulations. Given our speed and flexibility, we investigate a range of ionizing source and active quasar prescriptions. Spanning a large area of parameter space, we find order-of-magnitude fluctuations in the He~II ionization rate in the post-reionization regime. During reionization, the fluctuations are even stronger and develop a bimodal distribution, in contrast to semi-analytic models and the hydrogen equivalent.  These distributions indicate a low-level ionizing background even at significant He~II fractions.
\end{abstract} 

\begin{keywords}
cosmology: theory -- intergalactic medium -- diffuse radiation
\end{keywords}

\section{Introduction}

Throughout most of cosmic history, the metagalactic ionizing background evolved smoothly. The reionization of hydrogen ($z \gtrsim 6$) and of helium (fully ionized at $z \sim 3$) were two major exceptions. During these phase transitions, the intergalactic medium (IGM) became mostly transparent to the relevant ionizing photons, allowing the high-energy radiation field to grow rapidly. Aside from its fundamental importance as a landmark event in cosmic history, the full reionization of helium has several important consequences for the IGM.  Since helium makes up $\sim\!24$ per cent of the baryonic matter by mass, its ionization state significantly affects the mean free path of photons above 54.4~eV, the energy required to fully ionize helium. Additionally, reionizing helium should significantly increase the temperature of the IGM (e.g., \citealt{Hui97, Furl08a, McQu09}, hereafter M09). This heating may also affect galaxy formation by increasing the pressure (and hence Jeans mass) of the gas, which may affect the star formation histories of galaxies \citep{Wyit07}. Determining the timing and duration of this epoch is, therefore, important for understanding the evolution of the IGM.

Because of its large ionization potential, He~II reionization is driven by high-energy sources: quasars and other active galactic nuclei. This process, therefore, offers an indirect probe of these populations and their distribution through space and even inside dark matter halos. For example, the metagalactic background above 54.4~eV can even be used to constrain detailed properties of the sources, such as their lifetimes and emission geometries \citep{Furl11}.

The study of helium reionization has some advantages over its hydrogen counterpart (see \citealt{Bark01} for a review) and may serve as an object lesson for our understanding of that earlier era. Unlike for the reionization of hydrogen, we have excellent data on the state of the IGM at $z \sim 3$, including the temperature and density distribution (see \citealt{Meik07} for a review). Furthermore, the main drivers of helium reionization are quasars (due to their hard spectra), which are much better understood than their counterparts that drive hydrogen reionization. In particular, we know the quasar luminosity function (QLF) (e.g., \citealt{Wolf03, Rich06, Mast12}) and quasar clustering properties (e.g., \citealt{Ross09,Shen09,Whit12}).

Given the rarity of these sources and the inhomogeneity of the IGM, the process of helium reionization is expected to be patchy. This patchiness has important implications for the evolution of the IGM, and it is one of the crucial elements that must be modeled correctly.  Many measurements make assumptions about the UV-background radiation and the temperature-density relation, ignoring the inhomogeneity and timing of He~II reionization. Quantifying the expected spread in He~II ionization states is crucial.

Recent and ongoing high-resolution measurements of the He~II Lyman-$\alpha$ (\lya) forest have started to test this picture. \citealt{Dixo09} showed that early data indicated a rapid decrease in the optical depth with cosmic time at $z \ga 2.7$, which may be indicative of He~II reionization, while \citet{Furl10} showed that the large fluctuations observed in the optical depth were most likely indicative of ongoing reionization at $z \sim 2.8$. Now, the Cosmic Origins Spectrograph (COS) on the \emph{Hubble Space Telescope} is providing an even more powerful probe of the the IGM at $2.3 \la z \la 3.7$. The He~II optical depth along the explored sightlines varies significantly at $z > 2.7$ and generally increases rapidly at higher redshift \citep{Shul10,Wors11, Syph12, Syph13}. Four lines of sight also show significant variation in the He~II Ly$\beta$ optical depth \citep{Syph11}. This large set of observations -- and our detailed knowledge of the source populations and IGM at $z \sim 3$ -- make a detailed study of the He~II reionization process very timely.

Since quasars are rare, the ionizing background should exhibit significant fluctuations even after reionization is complete  \citep{Fard98, Bolt06, Meik07, Furl08b}. As mentioned above, the He~II \Lya forest measurements provide some direct evidence for this behavior \citep{Furl10}. Additionally, radiative transfer through the clumpy IGM can induce additional fluctuations \citep{Mase05, Titt07}. During reionization, fluctuations in the UV radiation background are even greater, because some regions receive strong ionizing radiation while others remain singly ionized with no local illumination. 

During the past decade, the reionization of helium and its impact on the IGM has received increased theoretical attention through a variety of methods. Both large-scale structures (spanning $\sim\!1$~Gpc comoving scales), given the rarity of quasars and their clustering properties, and the small scales of gas physics play a significant role in fully understanding this epoch, but accounting for both simultaneously is computationally challenging. Since this epoch is observationally constrained in a way that hydrogen reionization is not, a viable model conforms to these observations, such as the IGM density and temperature.

Several studies employ various combinations of analytic, semi-analytic, and Monte Carlo methods (e.g., \citealt{Gles05, Furl08, Furl08b}) to approximate the morphology of ionized helium bubbles, heating of the IGM, He-ionizing background, etc. None of these methods can comprehensively include all relevant physics, especially spatial information like source clustering. Many numerical simulations of this epoch focus on scales $\leq 100^3$ comoving Mpc$^3$ \citep{Soka02, Pasc07, Meik12}. These simulations miss the large ionized bubbles expected during helium reionization and fail to include many sources. M09 present large (hundreds of Mpc) $N$-body simulations with cosmological radiative transfer as a post-processing step. Note that M09 consider two classes of quasar models and vary the timing of reionization independently.

One major modeling hurdle is the implementation of quasars. The aggregate, empirical properties of quasars are well constrained with the QLF measured over a range of redshifts and to low optical luminosities (see \citealt{Hopk07}). The spectral energy distribution is more uncertain and varies significantly from quasar to quasar (e.g., \citealt{Telf02, Scot04, Shul12}). Furthermore, the lifetime of quasars (e.g., \citealt{Kirk08, Kell10, Furl11}) and their relation to dark matter halos (e.g., \citealt{Hopk06, Wyit07, Conr13}) are far from settled. The challenge is to satisfy the observed quantities while allowing for a range of possibilities for the more uncertain aspects.

To complement these numeric and analytic studies, here we present fast, semi-numeric methods that incorporate realistic source geometries to explore a large parameter space. We adapt the established hydrogen reionization code \textsc{DexM} \citep{Mesi07} to the $z = 3$ era, as appropriate for helium. The code applies approximate but efficient methods to produce dark matter halo distributions.  Using analytic arguments, ionization maps are generated from these distributions. The advantages of this approach are speed, as compared to cosmological simulations, and reasonably accurate spatial information (like halo clustering and a detailed, local density field), as compared to more analytic studies. We investigate a wide range of quasar properties.

These simulations are well-suited to study many features of the helium reionization epoch, including interpreting the He~II \Lya forest and estimating the He-ionizing background. In this paper, we consider the spatial morphology of He~II reionization and the He~II photoionization rate during and after reionization. Given our ability to span a large parameter space, we aim to pinpoint the assumptions that most strongly impact the results and to evaluate the effect of various uncertainties.

We use semi-numeric methods, outlined in \S\ref{sec:DexM}, to approximate the ionization morphology, density field, and sources of ionizing photons at $z = 3$. As a second layer, we place empirically determined active quasars in dark matter halos, following several prescriptions in \S\ref{sec:QSO}. With these inputs, we determine the He~II photoionization rate for post-reionization and several He~II fractions in \S\ref{sec:Gamma}. In \S\ref{sec:hydrogen}, we compare our results and methods with those for hydrogen reionization. We conclude in \S\ref{sec:disc}.

In our calculations, we assume a cosmology with $\Omega_m = 0.26, \Omega_{\Lambda} = 0.74, \Omega_b = 0.044, H_0 = 100h$~km s$^{-1}$ Mpc$^{-1}$) (with $h$ = 0.74), $n = 0.95$, and $\sigma_8 = 0.8$. Unless otherwise noted all distance quoted are in comoving units.

\section{Semi-Numeric Simulations of He~II Reionization} \label{sec:DexM}

To fully describe helium reionization and compute the detailed features of the \Lya forest, complex hydrodynamical simulations of the IGM, including radiative transfer effects and an inhomogeneous background, are required. Recent $N$-body simulations (e.g., M09) have advanced both in scale and in the inclusion of relevant physical processes. However, the major limitation is the enormous dynamic range, balancing large ionized regions with small-scale physics, required to model reionization. Generally, this problem is solved in two ways. First, most codes follow only the dark matter, assuming that the baryonic component traces it according to some simplified prescription.  They then perform radiative transfer on the dark matter field (perhaps with some modifications to reflect baryonic smoothing). Second, these codes add in the ionizing sources through post-processing, rather than through a self-consistent treatment of galaxy and black hole formation, which is not substantially better than simple analytic models. Also importantly, large simulations remain computationally intensive, limiting the range of parameter space that can be explored in a reasonable amount of time.

We adapt the semi-numeric code \textsc{DexM}\footnote{http://homepage.sns.it/mesinger/Sim.html} \citep{Mesi07}, originally designed to model hydrogen reionization at $z \ga 6$, for lower redshifts and for helium reionization. This code provides a relatively fast semi-numeric approximation to more complicated treatments but remains fairly accurate on moderate and large scales. For full details, see \citet{Mesi07}. In brief, we create a box with length $L = 250$~Mpc on each side, populated with dark matter halos determined by the density field. Given these halos, we generate a two-phase (singly and fully ionized) ionization field, where the He~III fraction is fixed by some efficiency parameter using a photon-counting method. For our high-level calculations, we additionally adjust the local density field to match recent hydrodynamical simulations.

There are drawbacks to our approach, of course.  For one, this semi-numeric method does not follow the progression of He~II reionization through a single realization of that process. In order to ensure accurate redshift evolution, photon conservation requires a sharp $k$-space filter, whereas we use a real-space, top-hat filter, applied to the linear density field (see the appendix of \citealt{Zahn07}; \citealt{Zahn11}). This filtering can result in ``ringing" artifacts from the wings of the filter response, in the limit of a rare source population (as is the case in He~II reionization). Moreover, during He~II reionization, the relevant ionizing sources are short-lived quasars.  When a source shuts off, the surrounding regions can begin to recombine in a complex fashion \citep{Furl08c}.  Our method does not explicitly follow these recombinations, as we will discuss below.  As a consequence, we perform all of our calculations at $z = 3$, near the expected midpoint of the He~II reionization according to current measurements. We expect our results to change only modestly throughout the redshift range $z \sim 2.5 - 3.5$, because the source density does not change significantly over this range and the IGM density field evolution is similarly limited.

Our primary concern in adapting \textsc{DexM} to this period, as compared to the higher redshift regime for which it has been carefully tested, is the increasing nonlinearity of the density field.  There are two possible problems. First, the dramatically larger halo abundance may lead to overlap issues with the halo-finding algorithm. Second, the IGM density field will have modest nonlinearities representing the cosmic web. We will address the latter problem in detail in \S\ref{sec:density}.

\subsection{Dark matter halos} \label{sec:halos}

The backbone of our method is the dark matter halo distribution. In a similar way to $N$-body codes, we generate a 3D Monte-Carlo realization of a linear density field on a box with $N = 2000^3$ grid cells ($L = 250$~Mpc). This density field is evolved using the standard Zel'dovich approximation \citep{ZelD70}. Dark matter halos are identified from the evolved density field using the standard excursion set procedure. Starting on the largest scale (the box size) and then decreasing, the density field is smoothed using a real-space top hat filter at each point. Once the smoothed density exceeds some mass-dependent critical value, the point is associated with a halo of the corresponding mass. Each point is assigned to the most massive halo possible, and halos cannot overlap. We choose our critical overdensity to match the halo mass function found in the \citet{Jenk01} simulations at $z = 3$, following the fitting procedure of \citet{Shet99}. The final halo locations are shifted from their initial (Lagrangian) positions to their evolved (Eulerian) positions using first-order perturbation theory \citep{ZelD70}. To optimize memory use, we resolve the velocity field used in this approximation onto a lower-resolution, 500$^3$ grid.

\begin{figure*}
   \vspace{-5\baselineskip}
 {
    \includegraphics[trim = 0 0 5.5cm 10cm, clip,width=0.48\textwidth]{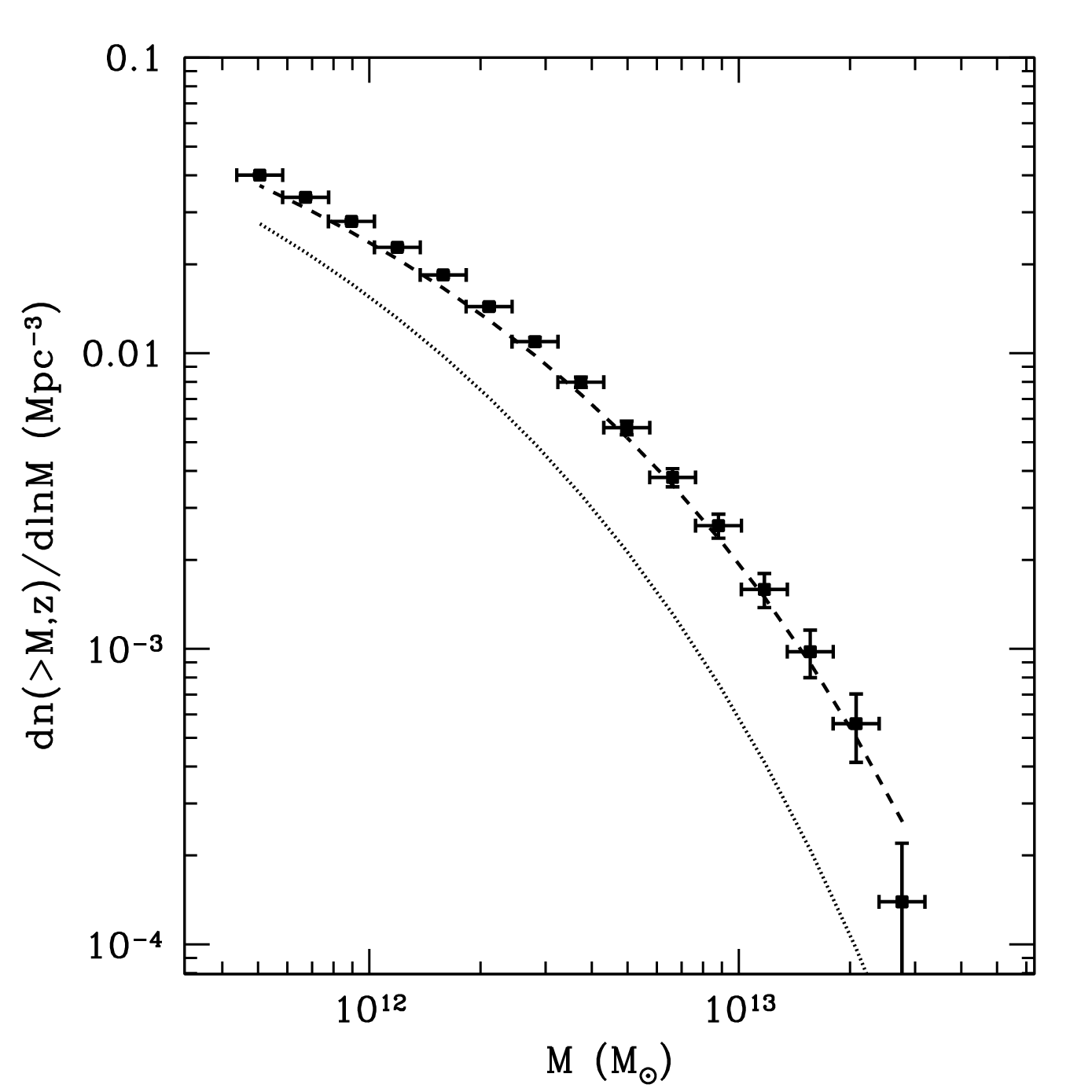}
   \includegraphics[trim = 0 0 5.5cm 10cm, clip,width=0.48\textwidth]{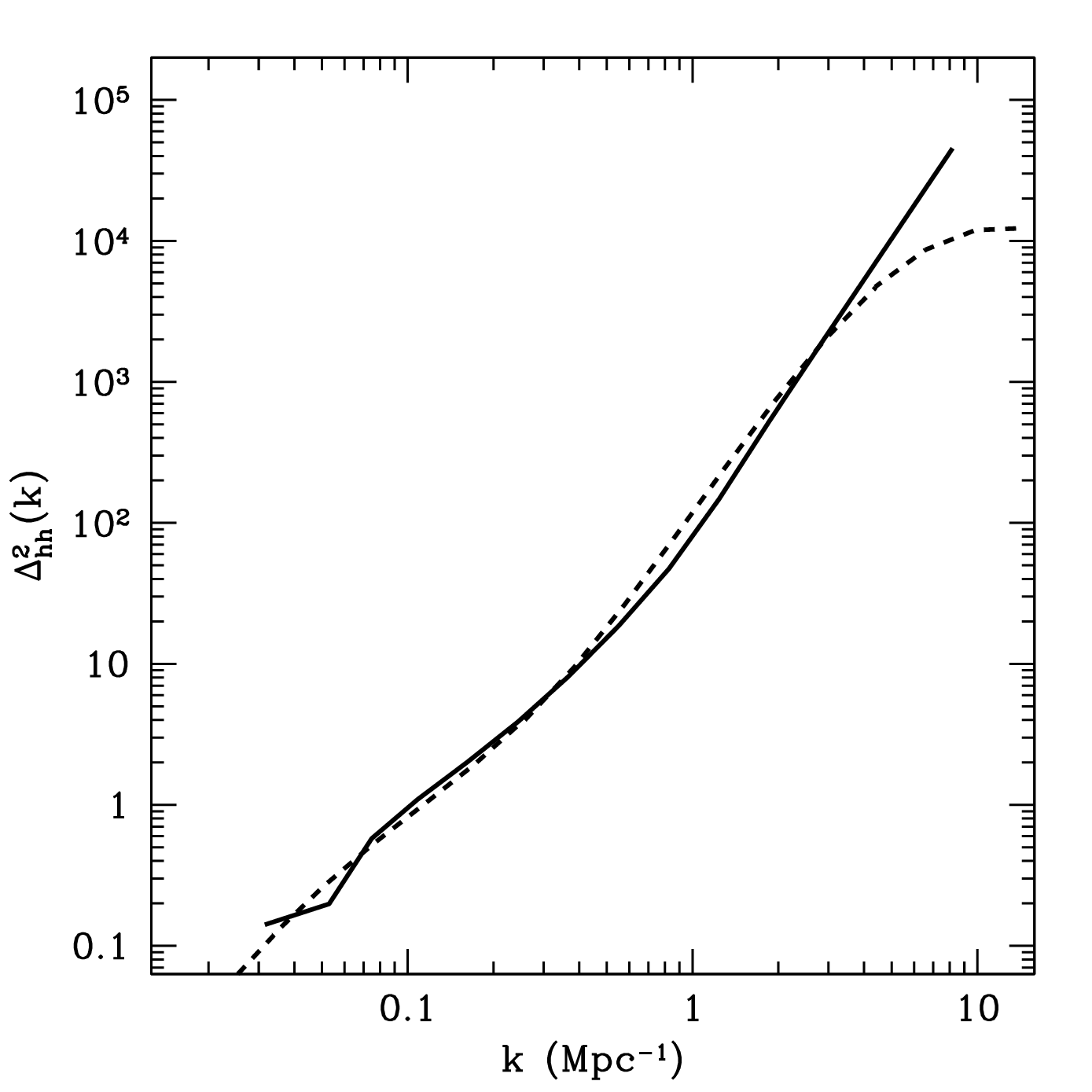} 
  }
   \caption{\emph{Left panel:} The calculated halo mass function shown as points. The dotted curve is the Press-Schecter analytic mass function. The dashed curve shows the Sheth-Tormen mass function, which agrees with our results by construction. \emph{Right panel:} Halo power spectra. The solid line represents the halo power spectrum measured from our standard ($L$ = 250~Mpc) box, and the dashed line is the halo model analytic equivalent.}
   \label{fig:mass_func}
   \vspace{-1\baselineskip}
\end{figure*}  

The left panel of Fig.~\ref{fig:mass_func} displays the resultant mass function as points, where the dashed line is the Sheth-Tormen \citep{Shet99} result with \citet{Jenk01} parameters and the dotted line is the Press-Schechter analytic solution for the mass function  \citep{Pres74}. We successfully reproduce the former mass function and, therefore, match large numerical simulations. The right panel shows our halo power spectrum, defined as $\Delta^2_{\rm hh}(k,z) = k^3/(2\pi^2V)\langle|\delta_{\rm hh} (\textbf{k},z)|^2\rangle_k$. Here, $\delta_{\rm hh} (\textbf{x},z) \equiv M_{\rm coll}(\textbf{x},z)/\langle M_{\rm coll}(z)\rangle - 1$ is the collapsed mass field, and the volume $V = L^3$. Also shown is the prediction from the halo model, which is a semi-analytic method for describing nonlinear gravitational clustering. This model matches many statistical properties of large simulations (see \citealt{Coor02}). Specifically, all dark matter is assumed to be bound in halos of varying sizes; then, given prescriptions for the distribution of dark matter within halos and the number and spatial distribution of halos, many large-scale statistical properties of dark matter can be calculated, including the halo power spectrum. At small scales, the measured power spectrum diverges from the halo-model equivalent, most likely because we do not include halo density profiles that would dominate at these scales.

Note that the halos with which we are concerned are the most massive at $z = 3$, containing only a few percent of the total mass. This fact means that overlap between mass filters is not very important, as it might be at lower masses: when a large fraction of the mass is contained inside collapsed objects, the halo filters could then overlap, in which case the ordering at which the cells were examined would affect the halo distribution and bias our results \citep{Bond96}. Fortunately, because we are concerned with quasars inside very massive dark matter halos, we avoid this issue. This focus on large halos also accounts for the large difference between the Sheth-Tormen and Press-Schechter mass functions in Fig.~\ref{fig:mass_func}. 

\subsection{Ionizing sources} \label{sec:source}

Because we focus on helium reionization, quasars are the relevant ionizing source.\footnote{Although some stars can produce He-ionizing photons, their contribution to reionization is most likely extremely small \citep{Furl08}.} Empirically, quasar properties are reasonably well constrained: the quasar luminosity function (QLF) has been measured to low optical luminosities over a range of redshifts (see the compilation in \citealt{Hopk07}).  However, the spectral energy distribution -- and hence the transformation of this observed luminosity function to the energies of interest to us -- is more uncertain. Furthermore, the relation between quasars and their host dark matter halos is not settled (e.g., \citealt{Hopk06, Wyit07, Conr13}). These details affect the ionization maps, the photoionization rate, and the pertinent range of the halo mass function via the minimum mass. Given these factors, there is no standard approach to the inclusion of quasars in numeric and analytic models. To address this uncertain relationship, we considered a wide range of models that encompass the spectrum of popular theories.

Given the distribution of dark matter halos found in the previous section, only the halos that host or have hosted quasars contribute to the ionization field. We begin by assuming that each halo above a minimum mass $M_{\rm min}$ is eligible to host a quasar and that a fraction $f_{\rm host} \le 1$ actually have hosted one (at either the present time or at some time in the past). Thus, these are the relevant sources for the ionization field calculation. As a fiducial value, we take $M_{\rm min} = 5\times 10^{11}$~\Msun. Next, we assume that these hosts produce $\zeta f_q$ He~II-ionizing photons per helium atom, where we have split this quantity into two factors for conceptional convenience. The first ($\zeta$) is the \emph{average} number of ionizing photons (minus recombinations, see below) across the entire source population, while we will use the latter ($f_q$) to describe how these photons are distributed amongst the hosts (i.e., it may vary with halo mass). If we then consider a region in which all the ionizing photons are produced by interior sources (and inside of which all of those photons are absorbed), the ionized fraction $x_{\rm HeIII}$ is then
\begin{equation} 
\label{eq:basic_one} 
\zeta f_{\rm host} = x_{\rm HeIII},
\end{equation}
where the $f_q$ factor has vanished because we are only considering the \emph{total} population (summed over all masses, so that the ionizing efficiency is simply the average value $\zeta$).

The efficiency parameter $\zeta$ can be estimated from first principles given a model for quasars (see, e.g., \citealt{Furl08}). One could compute the average number of He~II ionizations per baryon inside of quasars ($N_{\rm ion}$), estimate the efficiency with which baryons are incorporated into the source black holes ($f_{\rm BH}$), and estimate the fraction of these photons that escape their host ($f_{\rm esc}$); then $\zeta \sim f_{\rm BH} f_{\rm esc} N_{\rm ion}$, with additional corrections for the composition of the IGM and the number of photons ``wasted" through recombinations.  However, we note that $\zeta$ \emph{cannot} be inferred directly from the observed luminosity function, because it depends on the total number of ionizing photons emitted by a halo rather than the current luminosity. At minimum, we would require knowledge of the lifetime of the luminous phase of the quasar emission. Conversely, this implies that our choices for $M_{\rm min}$ and $f_{\rm host}$ enforce a quasar lifetime, which we will consider in \S\ref{sec:QSO}.

This $\zeta$ also depends on the recombination history of the IGM, as (to the extent that they are uniform) such recombinations essentially cancel out some of the earlier ionizations \citep{Furl04}. In practice, recombinations are inhomogeneous and so much more complex to incorporate \citep{Furl05}, especially because the small number of quasar sources can allow some regions to recombine substantially before being ionized again by a new quasar \citep{Furl08c}.  Our $f_{\rm host}$ parameter can approximately account for the latter effect (if one thinks of some of the ``off" halos as having hosted quasars so long in the past that their bubbles have recombined). We also expect that inhomogeneous recombinations are relatively unimportant for the quasar case, at least until the late stages of reionization \citep{Furl08,Davi12}. 

Because we lack a strong motivation for a particular value of $\zeta$, we will generally use it as a normalization constant. That is, we will specify \xHeIII~and a model for the host properties ($M_{\rm min}$ and $f_{\rm host}$), and then choose $\zeta$ to make eq.~(\ref{eq:basic_one}) true.  As described above, any such choice can be made consistent with the luminosity function if we allow the quasar lifetime to vary. 

Throughout most of the paper, we consider two basic source models. Neither is meant to be a detailed representation of reality: instead, they take crude, but contrasting, physical pictures meant to bracket the real possibilities. In the first, our \emph{fiducial method}, we imagine that some quasars existed long enough ago that any He~III ions they created have since recombined, so only a fraction of the halos (which recently hosted quasars) contribute to the ionization field.  We therefore assume that during reionization (1) only a fraction $f_{\rm host}$ of halos above $M_{\rm min} = 5 \times 10^{11}~\Msun$ have ever hosted sources, and that (2) each halo with a source, regardless of its mass, produces the same total number of ionizing photons. The latter assumption effectively fixes our $f_q$ parameter, which must vary inversely with mass.

We are left with two free parameters, $f_{\rm host}$ and $\zeta$.  For the first, we set $f_{\rm host} \approx x_{\rm HeIII}/Q_{\rm ion}$ as the fraction of halos that contain sources, where $Q_{\rm ion}$ is the (assumed) \emph{total} number of ionizing photons produced per helium atom (integrated over all past times).  Physically, this means that at the moment the universe becomes fully ionized we would have $f_{\rm host} \approx 1/Q_{\rm ion}$.  In this case, the other halos would have had ionizing sources so far in the past that their He~III regions have recombined since formation. Since we loosely assume $z = 3$ to be the midpoint of helium reionization, we then fix $\zeta$ so as to produce \xHeIII~= 0.5 with $f_{\rm host} = 0.167$, where $Q_{\rm ion} = 3$ as expected at $z = 3$. This efficiency effectively determines the size of each ionized bubble: thus, their number density and size together fix the net ionized fraction. To set other ionized fractions, $f_{\rm host}$ is varied, where intuitively more sources produce more ionizations.

In detail, these parameters cannot generally be set by analytic arguments, as the total ionized fraction depends on the numerical method as well (because our algorithm does not conserve or track individual photons). A final adjustment (of order unity) is necessary in order to reproduce the proper He~III fraction.  (Hence, $f_{\rm host} \approx x_{\rm HeIII}/Q_{\rm ion}$ above is not a strict equality.)  Thus, the physical parameters that motivate each realization should be taken only as rough guides to the underlying source populations.

Our second source model (which we will refer to as the \emph{abundant-source method}) is chosen primarily as a contrast to this fiducial one. We assume that (1) all halos above $M_{\rm min} = 5 \times 10^{11}~\Msun$ have hosted sources, even when the global \xHeIII~is small, and that (2) the ionizing efficiency of each halo is proportional to its mass (which demands that $f_q=1$). These assumptions mean that (when averaged over large scales) the quantity $\zeta f_{\rm host} f_{q}$ in eq.~(\ref{eq:basic_one}) reduces to the simple form $\zeta f_{\rm coll}$ (the same form typically used for hydrogen reionization; \citealt{Furl04}).  We then choose $\zeta$ to satisfy eq.~(\ref{eq:basic_one}) for a specified \xHeIII.

Neither of these approaches is entirely satisfactory, but they bracket the range of plausible models reasonably well and so provide a useful gauge of the robustness of our predictions. Our fiducial model fails to match naive expectations in two ways. First, it assumes that only a fraction of massive halos host black holes that have ionized their surroundings. One possible interpretation is that some of these sources appeared long ago, so that their ionized bubbles have recombined. However, in most cases the required number of recombinations would be much larger than expected \citep{Furl08c}. A simpler interpretation is that many of these halos simply lack black holes.  While this fails to match observations in the local Universe (see, e.g., \citealt{Gult09} and references therein), the situation at $z \sim 3$ is less clear, and it is possible that many massive halos have not yet formed their black holes.

On the other hand, this procedure does allow our calculations to crudely describe the evolution of the He~III field with redshift, even though the calculations are all performed at $z=3$.  The primary difference between our density field and one at, say $z=4$, is the decreased halo abundance in the latter.  Assuming that $f_{\rm host} < 1$, therefore, roughly approximates the halo field at this higher redshift (though, of course, it does not properly reproduce their spatial correlations, etc.).

A second problem with our fiducial model is that, by assuming that each halo produces the \emph{same} number of ionizing photons, the resulting black holes will likely not reproduce the well-known $M$--$\sigma$ relation between halo and black hole properties \citep{Gult09}. Our choice is motivated by two factors. First, any scatter in the high-redshift black hole mass-halo mass relation (as is likely if these halos are still assembling rapidly) may dominate the steep halo mass distribution. Second, some models predict a characteristic halo mass for quasar activity \citep{Hopk06}. In practice, this inconsistency is probably not an enormous problem, because our high mass threshold (and the resulting steep mass function shown in Fig.~\ref{fig:mass_func}) means that only a fairly narrow range of halos contribute substantially.

The abundant-source method at least qualitatively reproduces the relation between halo mass and black hole mass (and the expected ubiquity of quasars), but it leads to some other unexpected features. Most importantly, it demands that the ionized bubbles around many quasars be quite small when \xHeIII~$< 1$, as we will show later. Partly, this is because we work at $z=3$, where the number of massive halos is rather large.  At higher redshift, an identical prescription would find fewer sources, which would allow for larger individual bubbles.  Moreover, the characteristic luminosity of quasars does not vary dramatically throughout the helium reionization epoch, so we naturally expect that the average bubble size will also remain roughly constant (M09). 

Instead, in our calculations within this model, the assumed efficiency parameter $\zeta$ varies with \xHeIII~while the source distribution remains fixed. Thus, our sequence of ionized fractions should \emph{not} be viewed as a sequence over time in the abundant-source model, as we do not allow the halo field to evolve.  One could view the small sizes of the bubbles when \xHeIII~$< 1$ as a result of strong ``internal absorption" within the bubbles, which would keep them compact, but that does not seem a likely scenario \citep{Furl08}.

Nevertheless, it is useful to consider a contrasting case from our fiducial model, in which reionization proceeds not by generating more He~III regions but by growing those that do exist. Our abundant-source model provides just such a contrast and so is useful as a qualitative guide to the importance of the source prescription.

\subsection{Ionization field} \label{sec:ion_field}

With the ionizing source model and the dark matter halo distribution in hand, we determine which areas of our box are fully ionized and which are not. To find the He~III field, \textsc{DexM} uses a filtering procedure similar to that outlined \S\ref{sec:halos}. In particular, He~III regions are associated with large-scale overdensities in the ionizing source distribution. Thus, instead of comparing the evolved IGM density to some critical value as in \S\ref{sec:halos}, we consider the density of our sources (or, equivalently, of dark matter halos).   In the fiducial model, we let $N_{\rm host}({\bf x},R)$ be the total number of halos that have hosted sources inside a region of radius $R$ and mass $M$ centered at a point \textbf{x}. Then, this region is fully ionized if  \citep{Furl04}
\begin{equation} \label{eq:zeta_bub} \zeta \langle M_h \rangle N_{\rm host}(\textbf{x}, R) \geq M({\bf x},R), \end{equation}
where $\langle M_h \rangle$ is the average mass of halos containing sources.

First, the source field is smoothed to a lower resolution ($N_{\rm HeIII} = 500^3$ grid cells) to speed up the ionized bubble search procedure, which takes considerable time at high ionized fractions relative to the other components of the code. We then filter this source field using a real-space, top-hat filter, starting at a (mostly irrelevant) large scale $R_{\rm max}$ and decreasing down to the cell size. Every region satisfying eq.~(\ref{eq:zeta_bub}) at each filtering scale is flagged as a He~III bubble, regardless of overlap (unlike the halo-finding case). The result is a two-phase map of the ionization field, dependent on the number density of halos hosting ionizing sources.

\begin{figure*}
\vspace{+0\baselineskip}
{
\includegraphics[width=3.4cm]{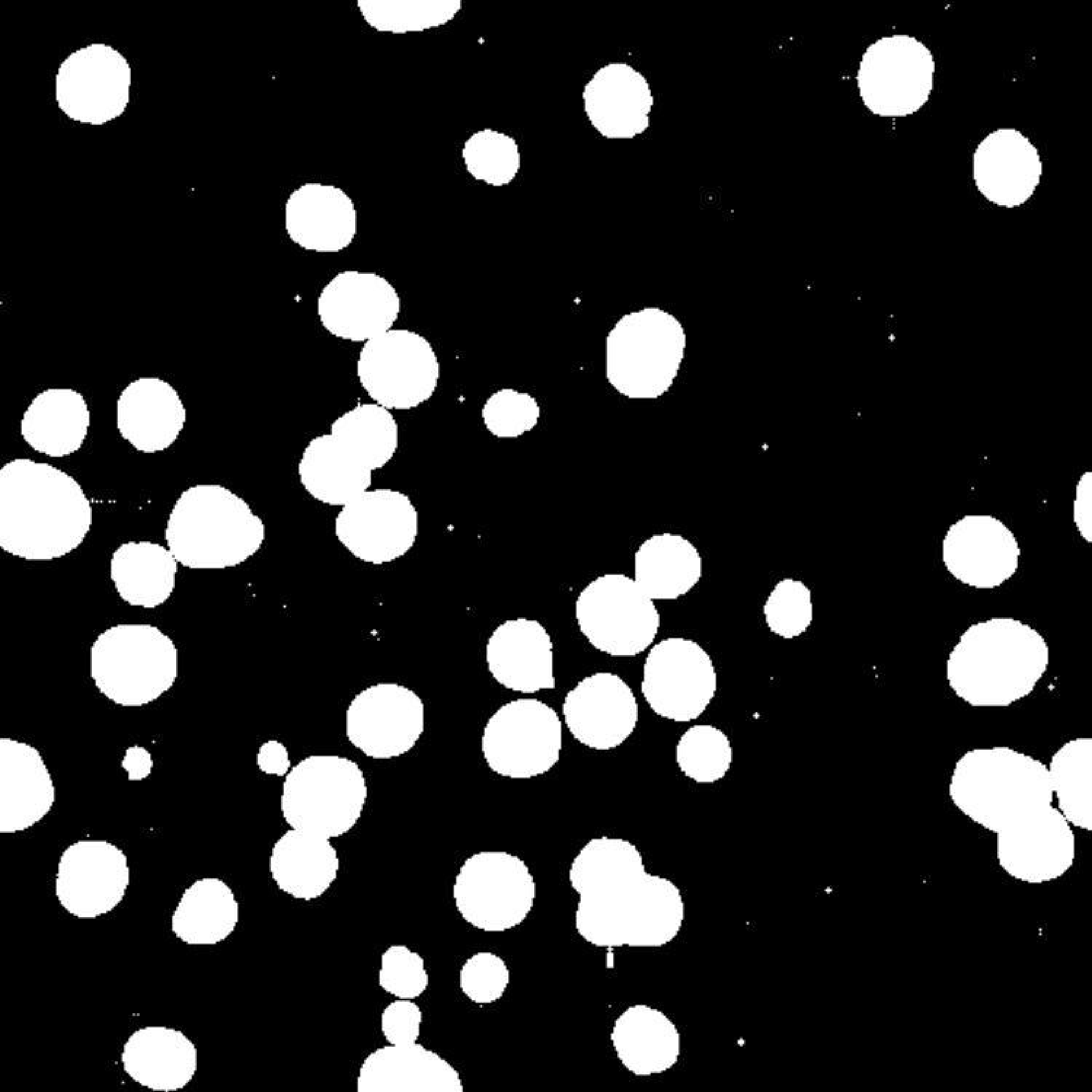}
\hspace{+8mm}
\includegraphics[width=3.4cm]{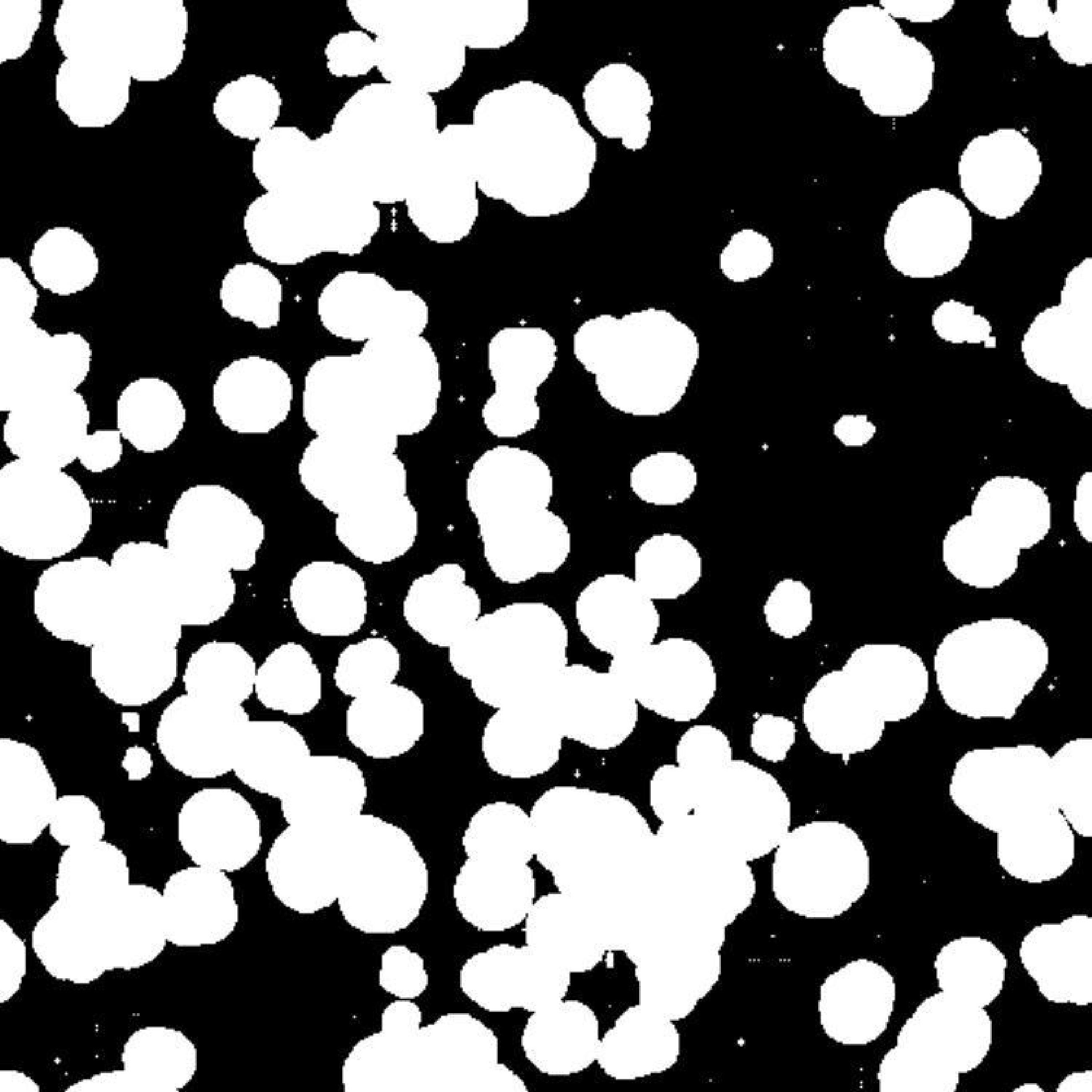}
\hspace{+8mm}
\includegraphics[width=3.4cm]{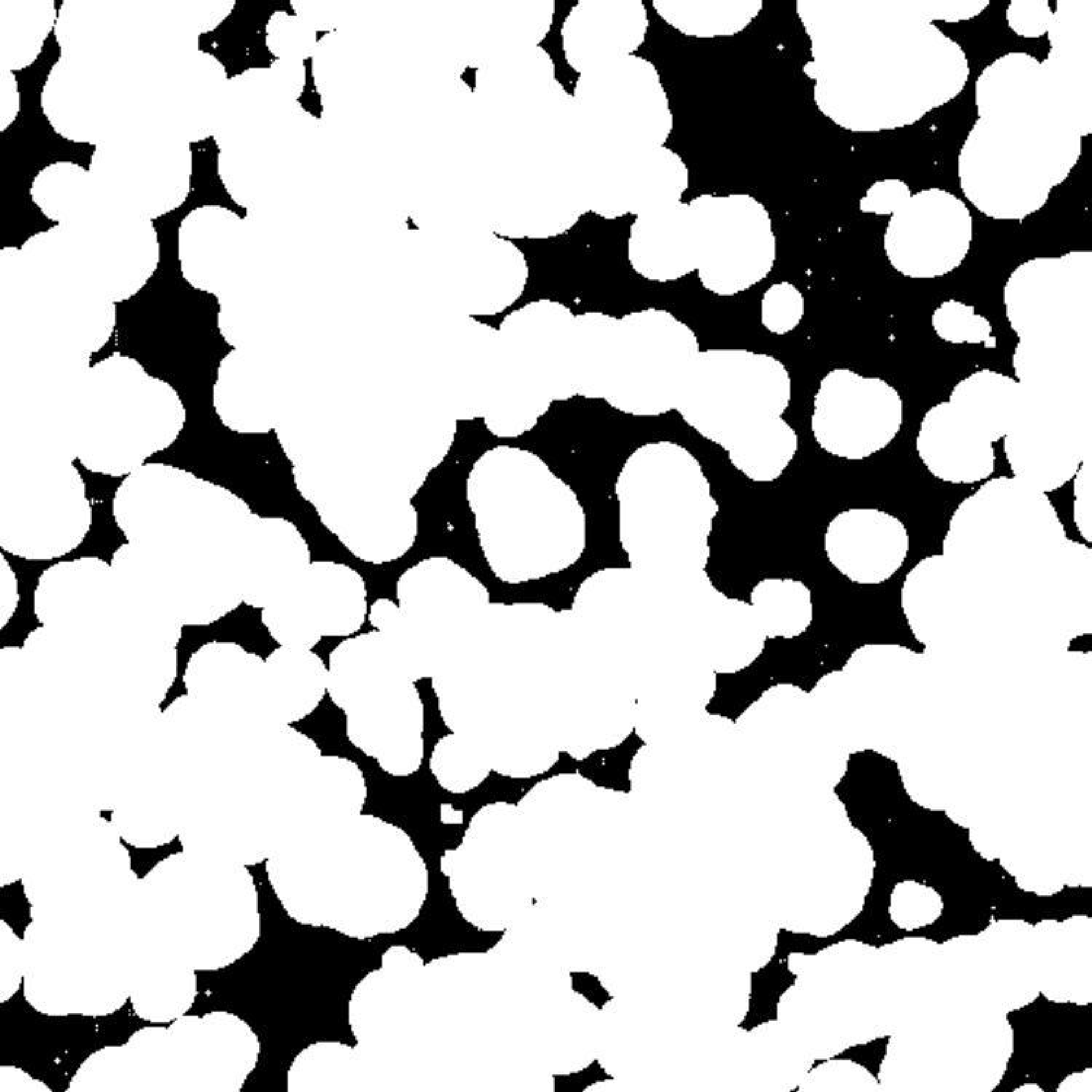}
\hspace{+8mm}
\includegraphics[width=3.4cm]{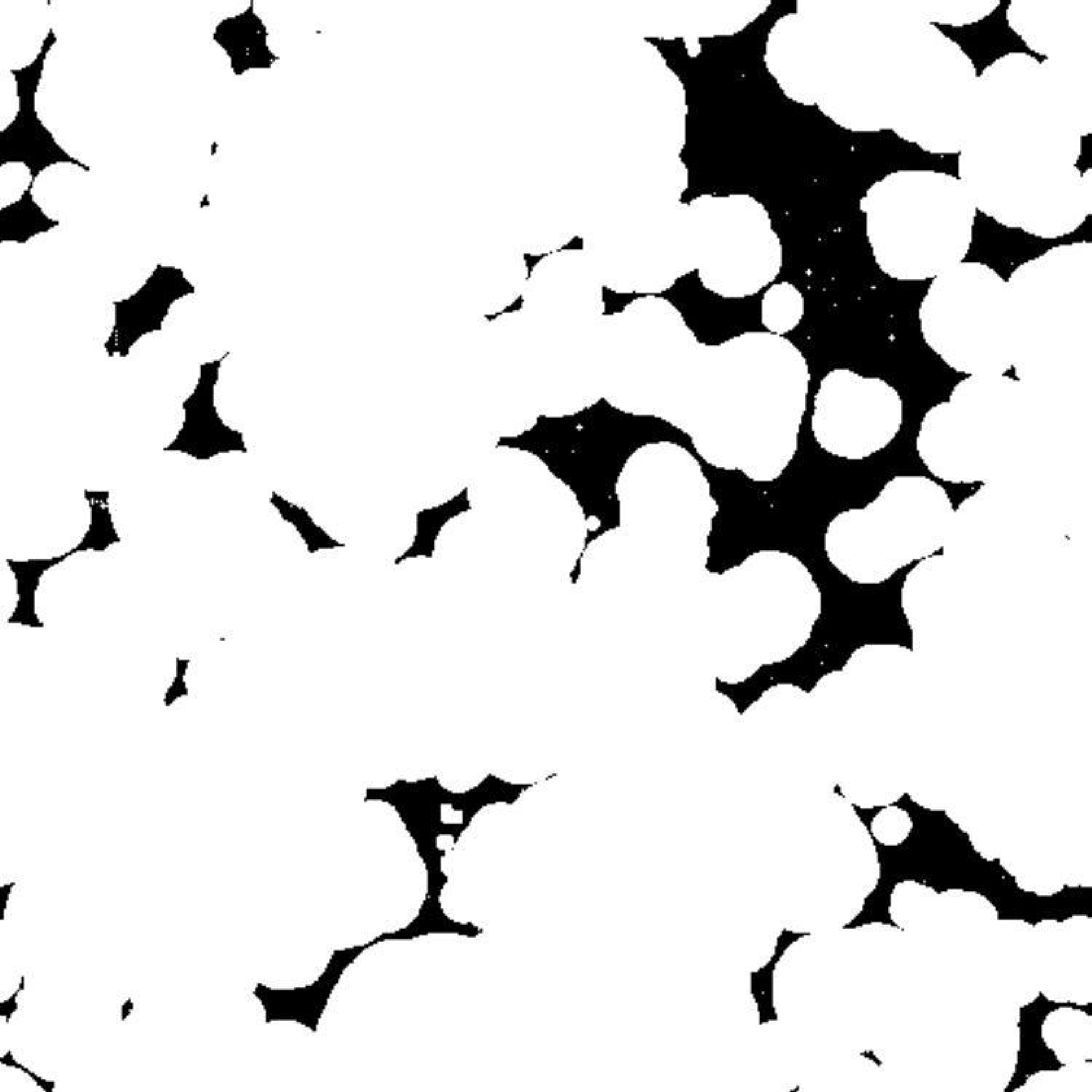}
}
\vspace{+0\baselineskip} \caption{
Slices from the ionization field at \xHeIII~= 0.20, 0.50, 0.80, and 0.90 (from left to right) for the fiducial model. All slices are 250~Mpc on a side and 0.5~Mpc deep. White (black) indicates He~III (He~II) regions.
\label{fig:bub_slices}
}
\vspace{-1\baselineskip}
\end{figure*}

The assumption of two phases with sharp edges is clearly a simplification. One-dimensional radiative transfer codes show that the ``transition region," or distance between fully ionized and singly ionized helium zones, is typically a few Mpc thick -- more precisely, the radius over which the He~III fraction falls from $\sim 0.9$ to $\sim 0.1$ is $\la 20$ per cent of the size of the fully ionized zone (F. Davies, private communication; see also fig. 19 of M09). The relatively late timing of He~II reionization, the rarity of the ionizing sources, and the hard spectra of quasars are responsible for widening this region. In many applications, especially those involving the IGM temperature, this transition region is very important. However, for other applications -- such as the He~II \Lya forest -- the two-phase approximation is adequate, because even a small He~II fraction renders a gas parcel nearly opaque. Fortunately, most He~III regions are very large (many tens of Mpc across), where many are built through the overlap of bubbles from individual sources, so the relatively narrow transition regions do not dominate the ionization topology. Nevertheless, we must bear in mind this important simplification when applying our semi-numeric simulations to observables.

In the ionization field slices shown in Fig.~\ref{fig:bub_slices}, we see the increasing ionized fraction through a static universe for the fiducial method. Displayed in the figure are \xHeIII~= 0.20, 0.50, 0.80, and 0.90, corresponding to $f_{\rm host} = 0.049, 0.167, 0.400,$ and 0.560. The map is fairly bubbly in nature with noticeable clustering. The fact that a pixel residing in a bubble has an increased likelihood of being near another bubble has important consequences for our later calculations. Even at the highest \xHeIII~displayed, there are large ``neutral" sections. Note the periodic boundary conditions. 

A similar procedure works for our abundant-source model, except the mass-weighting of the ionization efficiency per halo means that the relevant quantity is the total collapse fraction ($f_{\rm coll}(\textbf{x}, R)$).  In this case, the ionization criterion becomes very simple:
\begin{equation} \label{eq:zeta} f_{\rm coll}(\textbf{x}, R) \geq \zeta^{-1}. \end{equation} As mentioned in the previous section, this model results in small ionized bubbles that increase in size as \xHeIII~increases, with the highest fractions exhibiting very similar morphology to the fiducial model. Note that most large-scale applications should be fairly insensitive to the presence of small bubbles, since the radiation background is dominated by the brightest quasars within fully ionized regions.

\begin{figure*}
\vspace{+0\baselineskip}
{
\includegraphics[width=3.4cm]{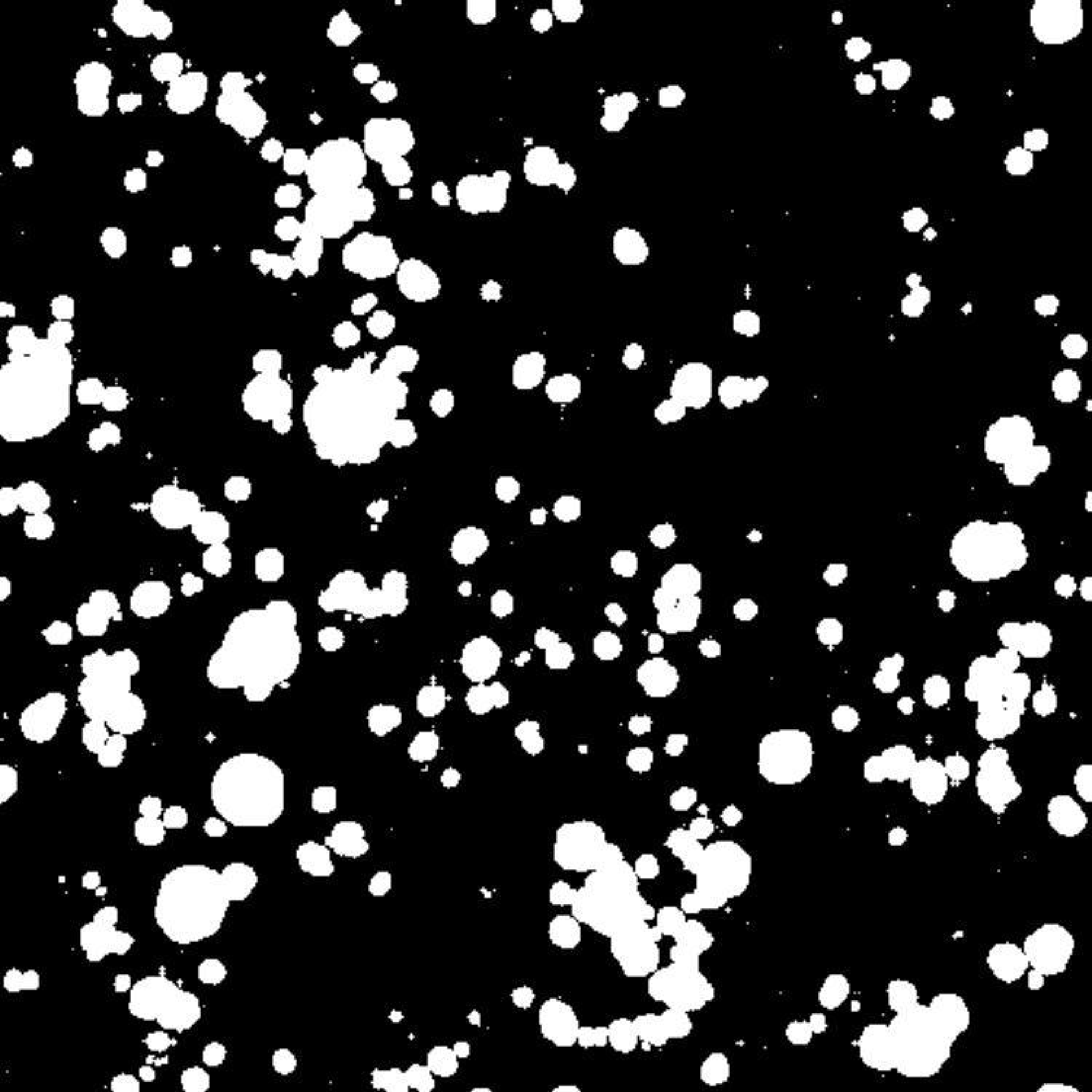}
\hspace{+8mm}
\includegraphics[width=3.4cm]{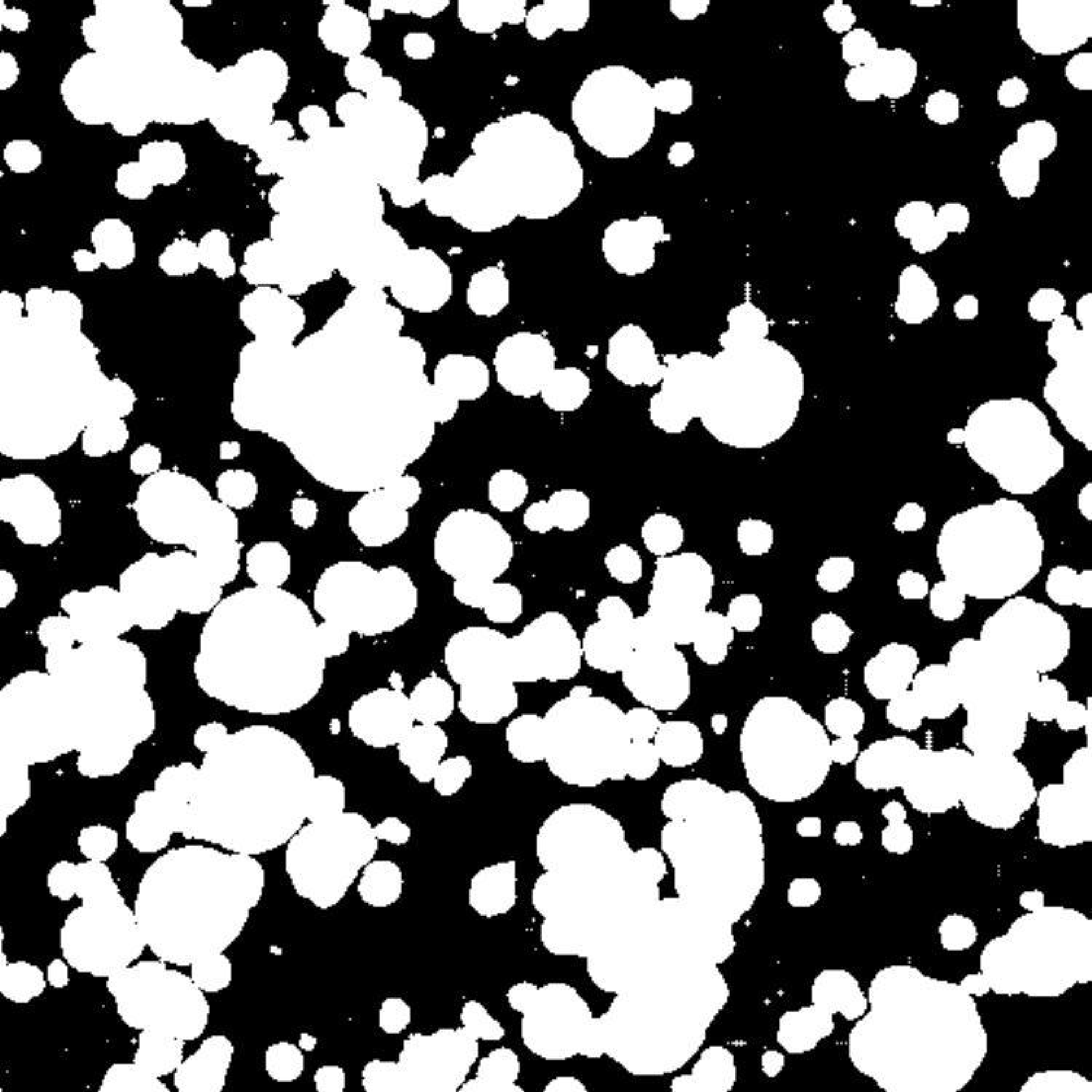}
\hspace{+8mm}
\includegraphics[width=3.4cm]{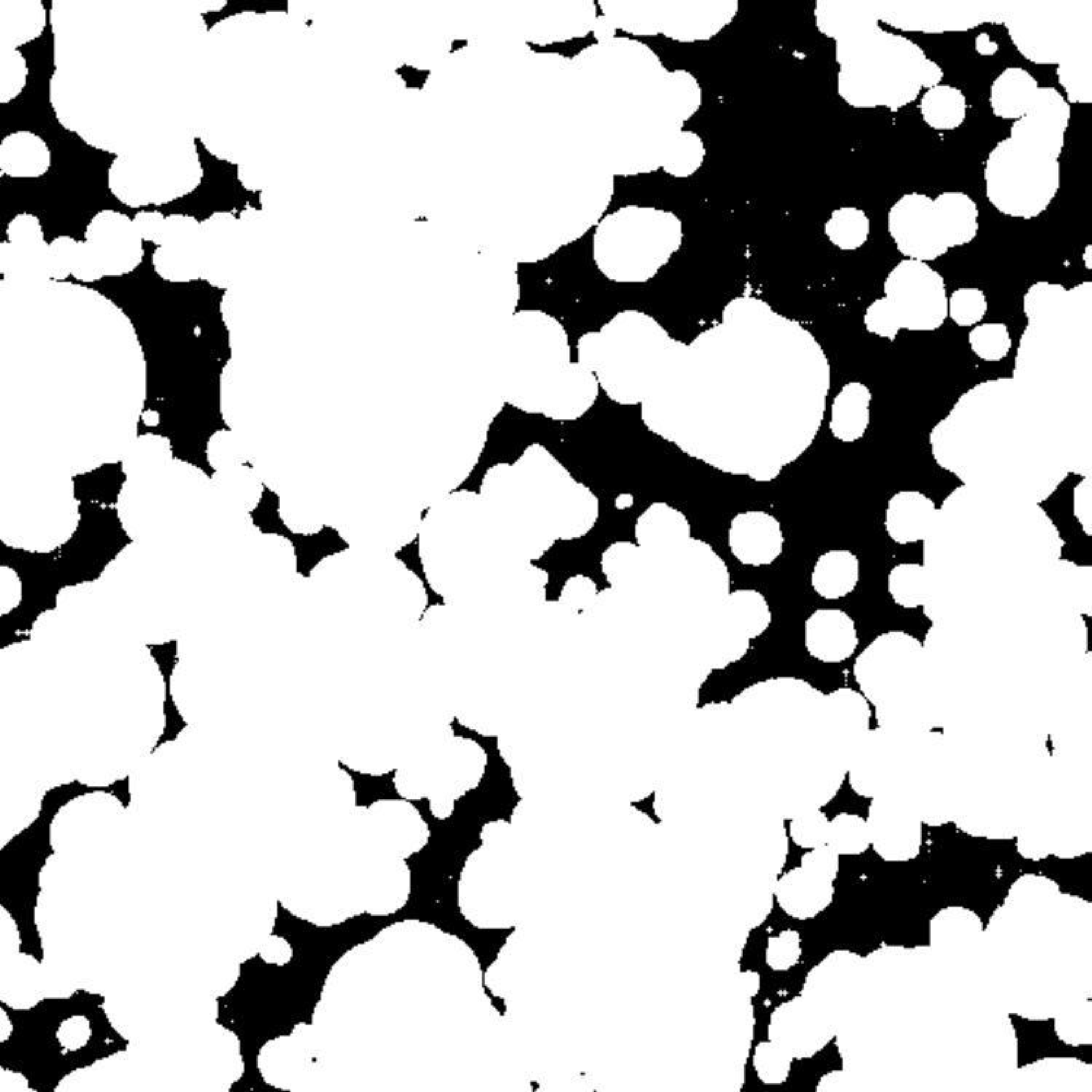}
\hspace{+8mm}
\includegraphics[width=3.4cm]{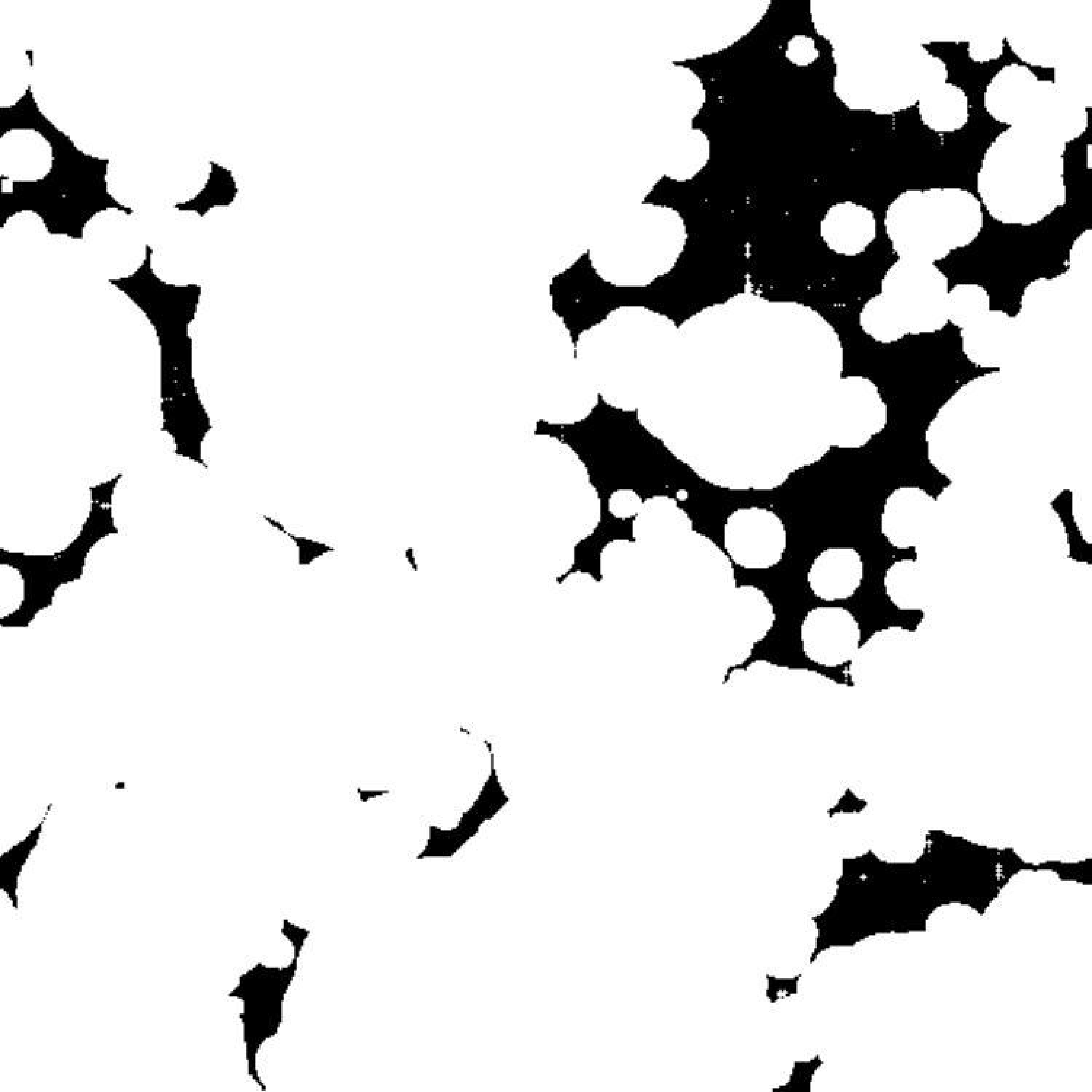}
}
\vspace{+0\baselineskip} \caption{
Slices from the ionization field at \xHeIII~= 0.21, 0.50, 0.80, and 0.90 (from left to right) for the abundant-source model. All slices are 250~Mpc on a side and 0.5~Mpc deep. White (black) indicates He~III (He~II) regions.
\label{fig:slices}
}
\vspace{-1\baselineskip}
\end{figure*}

\begin{figure*}
\vspace{+1\baselineskip}
{
\includegraphics[width=3.4cm]{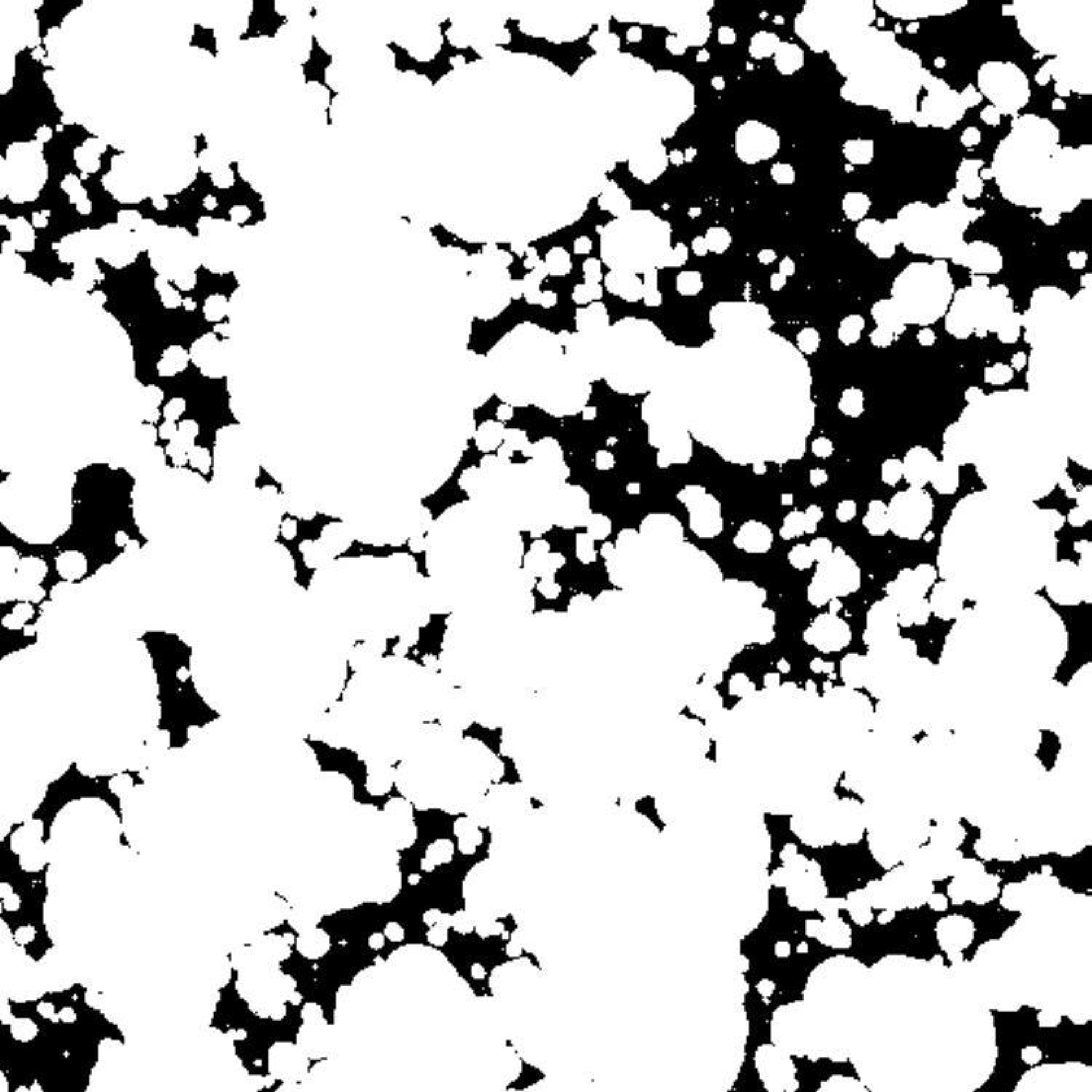}
\hspace{+8mm}
\includegraphics[width=3.4cm]{fig3c.eps}
\hspace{+8mm}
\includegraphics[width=3.4cm]{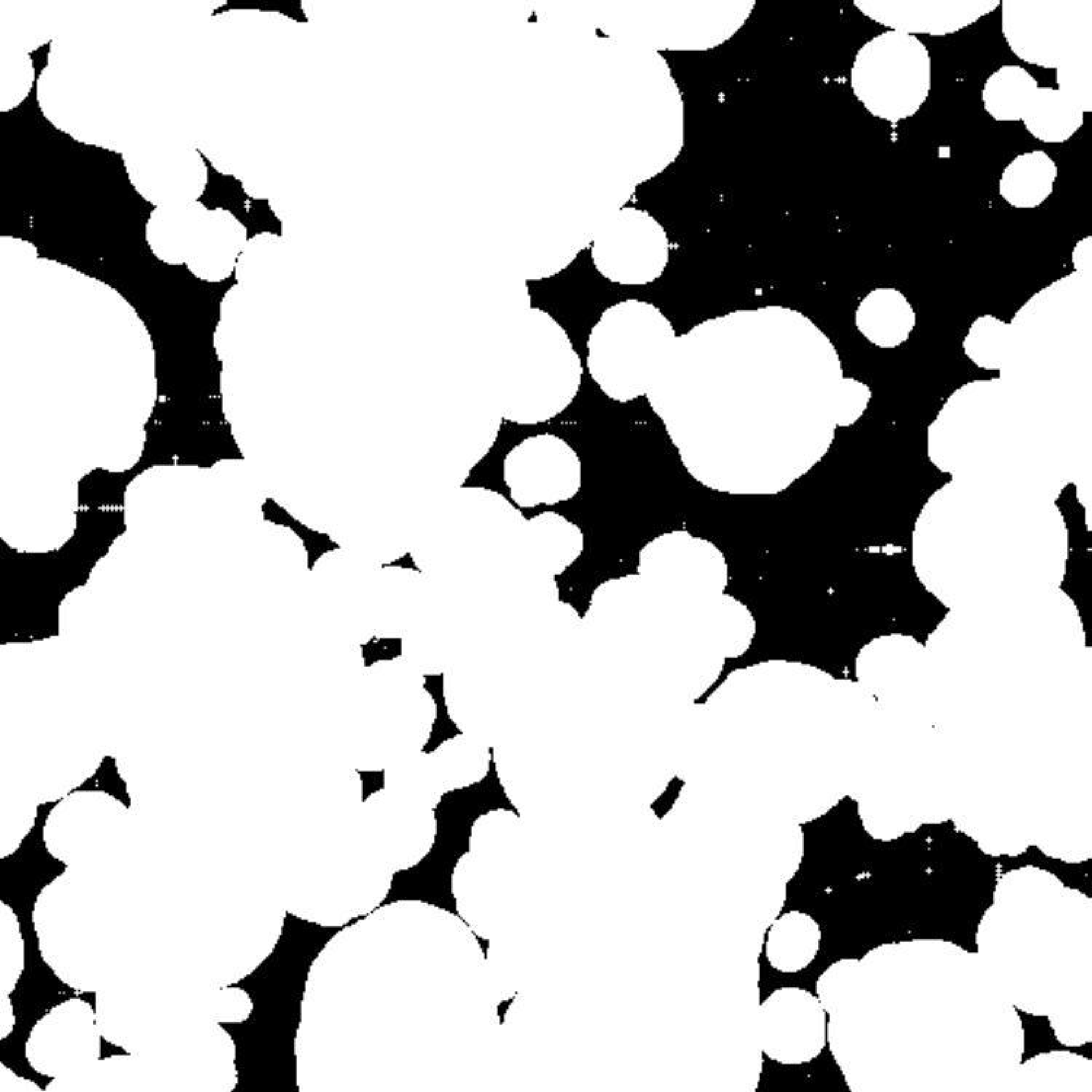}
\hspace{+8mm}
\includegraphics[width=3.4cm]{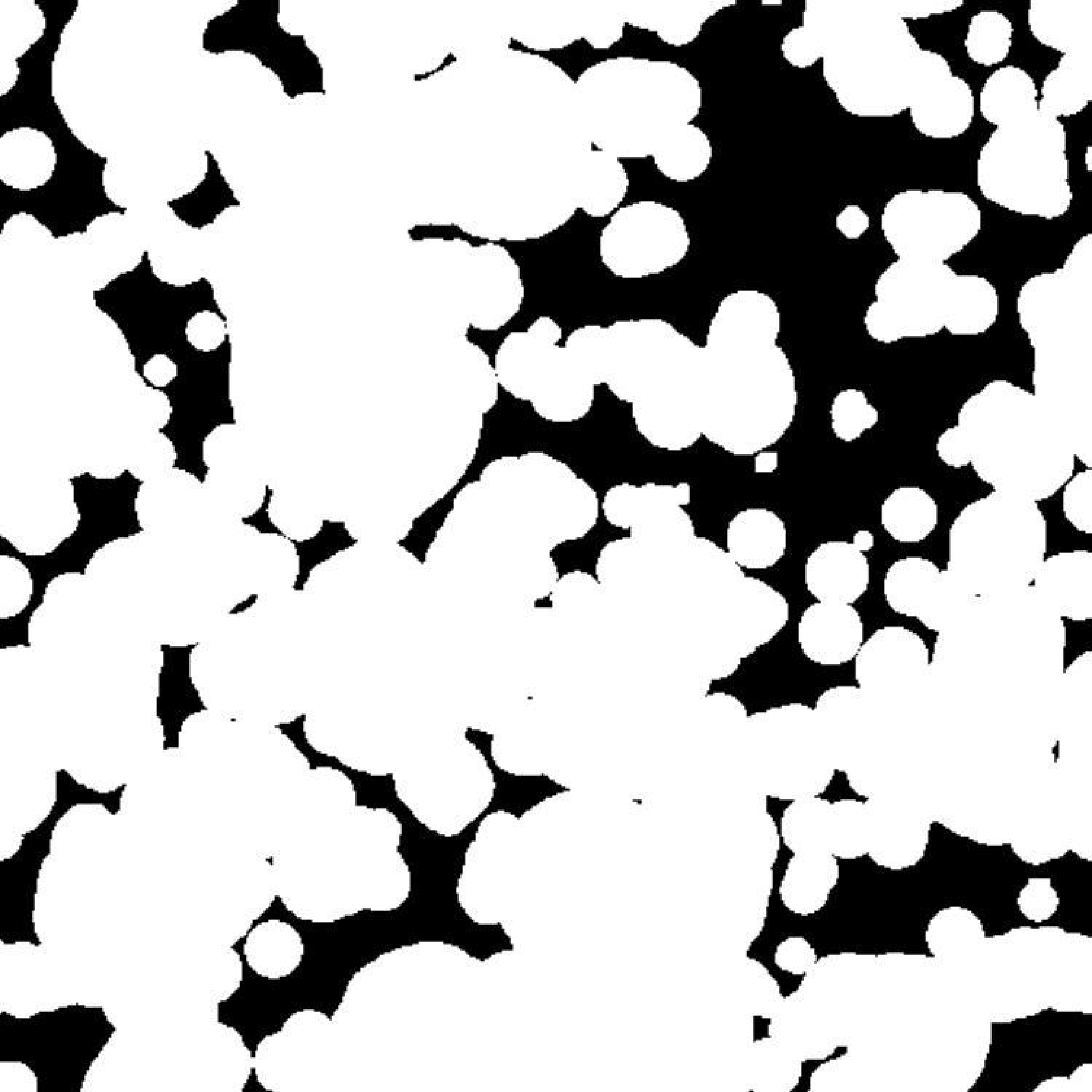}
}
\vspace{+0\baselineskip} \caption{
Slices from the ionization field for variations on the abundant-source model at \xHeIII~$\approx 0.80$. From left to right, $M_{\rm min} = (0.5, 5, 10)\times 10^{11}~\Msun$. The rightmost panel shows the fiducial model with $f_{\rm host} = 1$ at $M_{\rm min} = 5\times 10^{11}~\Msun$.
\label{fig:variation}
}
\vspace{-1\baselineskip}
\end{figure*}

Beyond our two main methods, we consider some additional parameter choices, shown in Fig.~\ref{fig:variation}. In the left three panels, we explore a range of minimum halo mass (effectively changing the quasar lifetime) for the abundant-source method. From left to right, $M_{\rm{min}} = (0.5, 5, 10)\times 10^{11}~\Msun$ at \xHeIII~$\approx 0.80$. As $M_{\rm{min}}$ is decreased, the number of sources increases, which implies that the quasar is ``turned on" for a shorter period of time, ionizing a smaller region given a fixed \xHeIII. Similarly, a higher minimum mass means fewer sources and larger, rarer ionized bubbles, which is more pronounced at lower \xHeIII. The rightmost panel of Fig.~\ref{fig:variation} shows the result for the fiducial method with $f_{\rm{host}} = 1$ and \xHeIII~$\sim 0.80$. Importantly, the second panel (abundant-source model with $M_{\rm{min}} = 5\times 10^{11}~\Msun$) and this panel are very similar visually, indicating the exact ionization criterion is unimportant.

\begin{figure*}
   \vspace{-5\baselineskip}
 {
    \includegraphics[trim = 0 0 5.5cm 10cm, clip,width=0.48\textwidth]{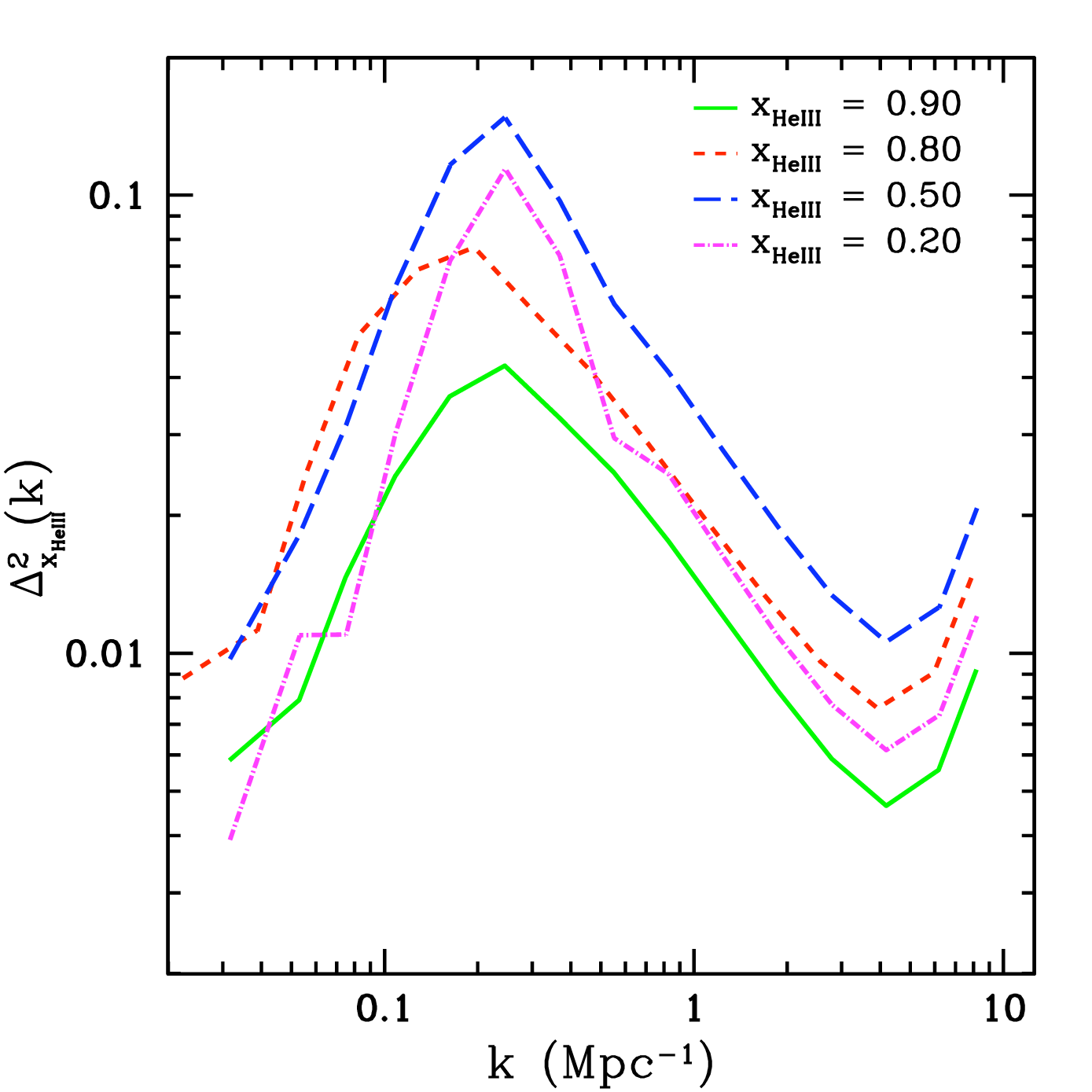}
   \includegraphics[trim = 0 0 5.5cm 10cm, clip,width=0.48\textwidth]{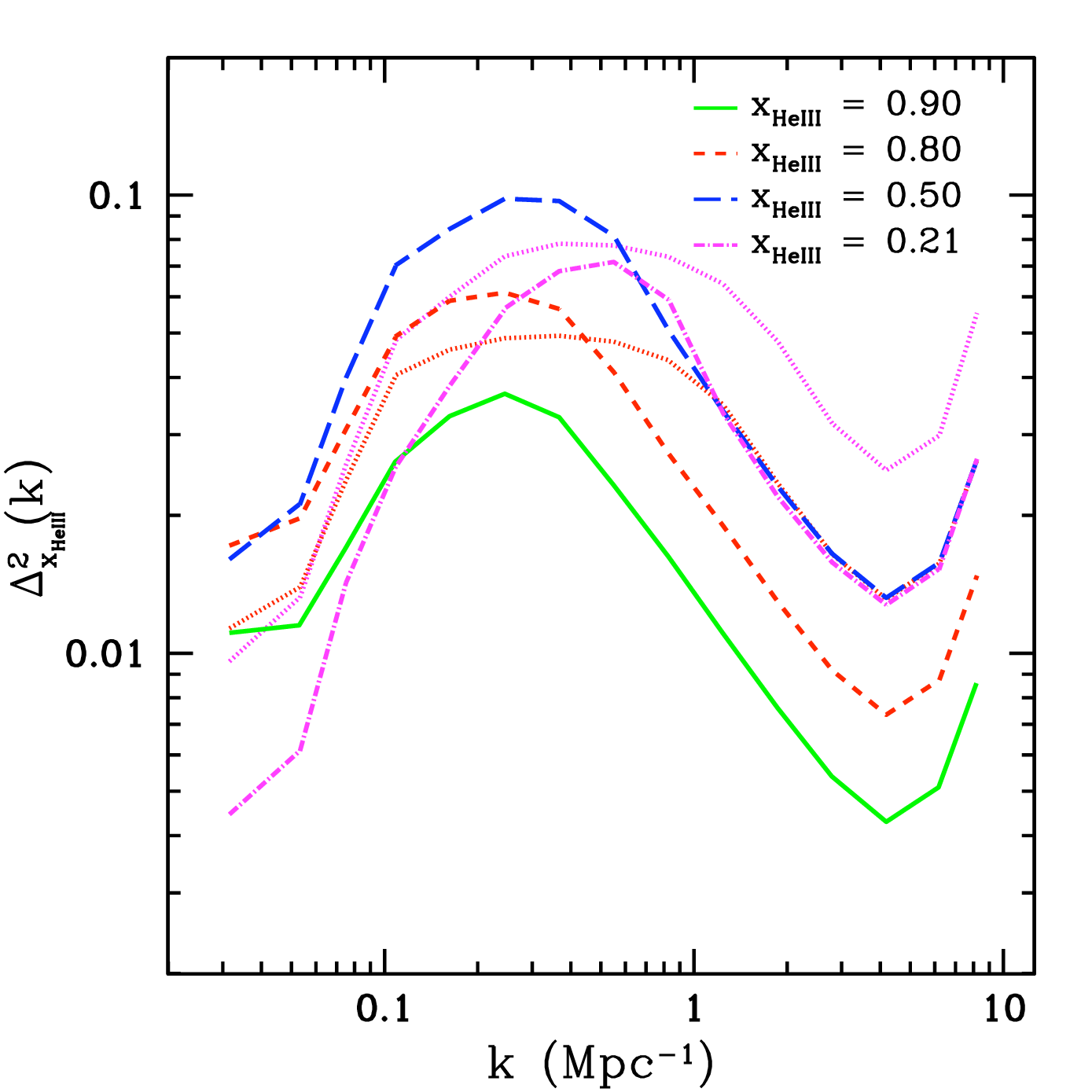} 
  }
   \caption{ The power spectrum of \xHeIII~at various ionized fractions. \emph{Left panel:} Shown are \xHeIII~= 0.20 (dot-dashed), 0.50 (long-dashed), 0.80 (short-dashed), and 0.90 (solid) for the fiducial model. \emph{Right panel:} The abundant-source method for \xHeIII~= 0.21 (dot-dashed), 0.50 (long-dashed), 0.80 (short-dashed), and 0.90 (solid) is plotted. The dotted lines represent the inclusion of even more sources ($M_{\rm min} = 0.5\times 10^{11}~\Msun$) for \xHeIII~= 0.50 and 0.80. }
    \label{fig:ion_ps}
   \vspace{-1\baselineskip}
\end{figure*}

To quantify the statistical properties of our ionization field, we calculate the dimensionless ionized power spectrum $\Delta^2_{x_{\rm HeIII}}(k) = k^3/(2\pi^2V) \langle |\delta_{x_{\rm HeIII}}(\mathbf{k})|^2\rangle _k$, where $\delta_{x_{\rm HeIII}}(\mathbf{x}) \equiv x_{\rm HeIII}(\mathbf{x})/\bar{x}_{\rm HeIII} - 1$. The results for the fiducial model at \xHeIII~= 0.20, 0.50, 0.80, and 0.90 are shown by the dot-dashed, long-dashed, short-dashed, and solid lines, respectively, in the left panel of Fig.~\ref{fig:ion_ps}. These curves are qualitatively similar to M09 results (except for the lowest \xHeIII, which we expect to differ the most), with a large-scale peak and small-scale trough. Differences in the underlying quasar models affect the exact position of the peak, but our results are comparable: $\sim\!30$~Mpc scales compared to $\sim\!60$~Mpc in M09. Note that the positions of the peaks and troughs do not vary with the He~III fraction, which is consistent with M09. Since, in this fiducial model, we are only changing the number of halos that contribute to the ionization at different \xHeIII, the size of the ionized bubbles remain constant, leading to the lack of evolution in the peak scale with \xHeIII. A similar phenomenon occurs in M09, because the typical luminosity of quasars remains roughly constant with redshift (as does the effective lifetime in their model), so the size of a typical ionized bubble remains the same throughout reionization. 

The right panel of Fig.~\ref{fig:ion_ps} shows the results for the abundant-source model with \xHeIII~= 0.21, 0.50, 0.80, and 0.90 (dot-dashed lines, long-dashed, short-dashed, and solid, respectively). The peaks are wider and evolve to smaller scales as \xHeIII~decreases, since the lower ionization states are dominated by smaller and smaller bubbles. As expected from the ionization morphologies, the lower \xHeIII~exhibit the largest differences from the fiducial model. The dotted lines assume the lowest minimum halo $M_{\rm min} = 0.5 \times 10^{11}~\Msun$ at \xHeIII~= 0.50 and 0.80, as shown in the leftmost panel of Fig.~\ref{fig:variation}. Since each ionized bubble is smaller than the $M_{\rm min} = 5 \times 10^{11}~\Msun$ case, the peak in the spectrum is shifted to the right and generally smoother. Omitted from this panel, the power spectra of the other variations outlined in Fig.~\ref{fig:variation} are nearly indistinguishable. 

\subsection{Density field} \label{sec:density}

We use the density field not just to find dark matter halos and ionized regions but also to track photon transfer through the IGM. Specifically, when calculating the photoionization rate for He~II later in this paper, we wish to follow the penetration of hard photons into neutral regions. This task requires a reasonably accurate small-scale density field appropriate to baryons. This density field is also crucial to investigating other observable quantities, such as the He~II \Lya forest. In this section, we describe our procedure for generating that distribution.

\begin{figure}
   \vspace{-5\baselineskip}
 {
    \includegraphics[trim = 0 0 5.5cm 10cm, clip,width=0.48\textwidth]{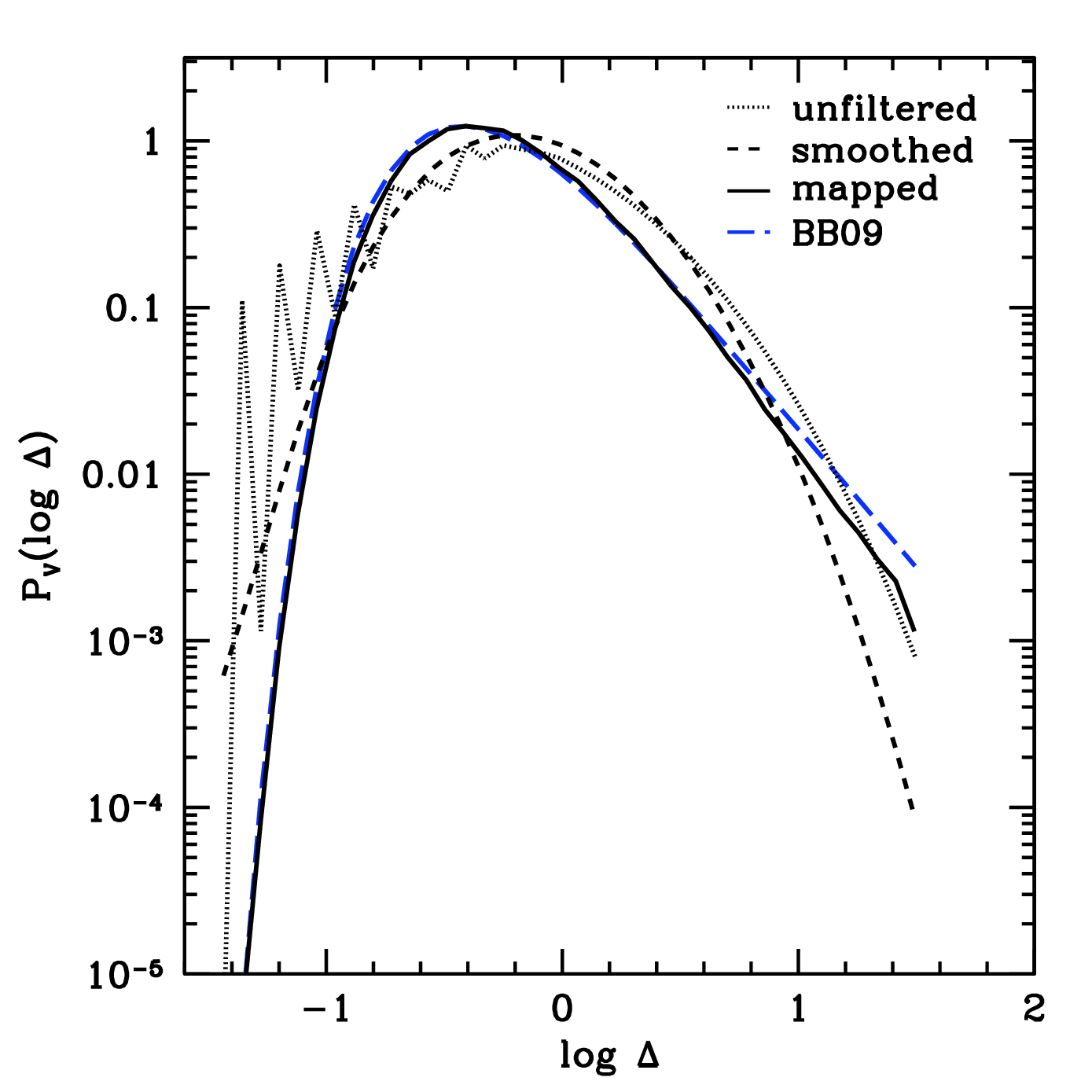}
  }
   \caption{The volume-weighted density distribution before, during, and after our mapping procedure with a reference curve. The dotted (short-dashed) line is the evolved (then smoothed) density distribution. The solid line shows the distribution used for subsequent calculations, which is mapped onto the BB09 fit, shown as the long-dashed line.}
   \label{fig:delta}
   \vspace{-1\baselineskip}
\end{figure}

As mentioned above, the density field in \textsc{DexM} is generated through first order perturbation theory.  Specifically, matter particles on the high-resolution (2000$^3$) grid are moved from their Lagrangian to their Eulerian coordinates using the displacement field \citep{ZelD70}, and then binned onto the lower-resolution (500$^3$) grid.  The resulting field has an inherent particle mass quantization and can be smoothed with a continuous filter to obtain a continuous density field (as in all particle dynamics codes).  Ideally, the filter scale should correspond to the Jeans mass. We therefore smooth the density field by applying a real-space, top-hat filter over 0.75~Mpc (or 1.5 pixel lengths). Note that the Jeans length in ionized gas at the mean density with an IGM temperature of $\sim\!10^4$~K is  $\sim\!1$ Mpc (near our smoothing size of 0.75~Mpc). The pre- and post- smoothing density PDFs are shown in  Fig.~\ref{fig:delta} with the dotted and dashed lines respectively.

This approach tends to underpredict overdensities and overpredict voids, although \citet{Mesi11} show that during hydrogen reionization ($z \gtrsim 6$), the resulting density field agrees with one generated from a hydrodynamical simulation at the percent level to about a dex around the mean density (spanning a large majority of the IGM).  However, at the redshifts of interest for He~II reionization, the density field becomes more nonlinear, and our first-order perturbation theory approach is less accurate.  Rather than adding higher order corrections, we apply a simple mapping procedure to the smoothed-particle hydrodynamics results of \citet[hereafter BB09]{Bolt09b}, using the respective cumulative probably distributions of the overdensity $p_{\rm DF}(<\Delta_{\rho})$ from our calculation and $p_{\rm BB}(<\Delta_{\rho})$, the fit from BB09. Here, $\Delta_{\rho} = \rho/\bar{\rho}$. The long-dashed curve in Fig.~\ref{fig:delta} is the BB09 fit. For each cell's overdensity $\Delta_{\rho}^i$, we assign a new overdensity $\Delta_{\rho}^j$ such that $p_{\rm DF}(<\Delta_{\rho}^i) = p_{\rm BB}(<\Delta_{\rho}^j)$. In this way, we preserve the ordering of our underlying matter distribution, but we also achieve a result closer to the gas distribution, which is more appropriate for computing observables.

The final, mapped density distribution is the solid curve in the figure. The discreteness visible in the dotted line and a lack of very large overdensities render the matching procedure imperfect. The choice of filter scale has minimal effect on our later calculations. Compared to the smoothed field without mapping, the mapped distribution slightly narrows the photoionization curves in \S\ref{sec:Gamma}, since higher density regions effectively absorb more photons. 

\section{Quasar models} \label{sec:QSO}

We are interested not only in how the source field affects the morphology of ionized gas but also in how it affects the metagalactic radiation background.  This background directly affects the \Lya forest and also indirectly affects several interesting observables, like the temperature of the IGM and ionization states of metal line absorbers. Although obviously intertwined with the determination of the ionization field, we must take an additional step to determine this background, which depends on the \emph{instantaneous} luminosities of the sources. Conversely, the ionization maps depend on the \emph{history} of ionizing photons. We follow an empirical approach to determine the number and intrinsic properties of the ``active" quasars, using an observationally determined QLF. Then, we compare three basic models for placing these quasars in dark matter halos.  

The \citet{Hopk07} QLF, $\Phi(L,z)$, serves as the starting point for all three models. First, we determine the number density of quasars at the redshift in question
\begin{equation} \label{eq:nQ} n_q ~~=~~ \int_{L_{\rm min}}^{L_{\rm max}} \Phi (L,z)~dL, \end{equation} where we take $L_{\rm min} = 10^{9.36}~L_\odot$ and $L_{\rm max} = 10^{17.05}~L_\odot$ in the $B$-band (4400~\AA). Our conclusions are extremely insensitive to these imposed cuts. At $z$ = 3, the comoving quasar density is then $n_q = 2.566\times10^{-5}$~Mpc$^{-3}$. Therefore, 400 quasars ($N_q$) reside in our volume. This quantity, together with the total number of halos ($N_{\rm halo}$), $f_{\rm host}$, and the Hubble time at $z = 3$ imply a quasar lifetime
\begin{equation} \label{tQSO} t_{\rm QSO} ~~=~~ \frac{N_q}{f_{\rm host}N_{\rm halo}}~\times~H^{-1}(z). \end{equation} The $t_{\rm QSO}$ ranges from approximately 1--10~Myr for our suite of models, consistent with recent estimates (e.g., \citealt{Kirk08, Kell10, Furl11}). Note how a smaller $f_{\rm host}$ increases the typical lifetime:  if the observed quasars are generated by a smaller set of black holes (because fewer halos host them), each one must live for longer.

We assign each quasar a $B$-band luminosity $L_B$, defined as $\nu L_{\nu}$ evaluated at 4400~\AA, randomly sampled from the QLF. Note that since we are concerned with He-ionizing radiation, this $B$-band luminosity must be converted to much higher frequencies. We assume a broken power-law spectral energy distribution \citep{Mada99}:
	\begin{equation} L_{\nu} \propto   \left\{ \begin{array}{ll}
	\nu^{-0.3}		& 	~~~2500 < \lambda < 4600~\mbox{\AA}\\
	\nu^{-0.8}		& 	~~~1050 < \lambda < 2500~\mbox{\AA}\\
	\nu^{-\alpha}	& 	~~~\lambda < 1050~\mbox{\AA}. \end{array} \right. \end{equation} A large spread exists in the measured extreme-UV spectral index $\alpha$ values for individual quasars. To replicate the \citet{Telf02} distribution for quasars, we model $\alpha$ as a gaussian distribution with mean value $\langle\alpha\rangle = 1.5$ and a variance of unity, constrained by $\alpha \in (0.5, 3.5)$. Each quasar is, therefore, given a random value for $\alpha$ within this distribution.

Now that each quasar has a designated specific luminosity, $L_{\nu}$ (given by $L_B$ and $\alpha$), the three quasar models differ in the method for placing them inside halos. Note that the two prescriptions for the ionization morphology only determine which host halos have ``turned on" by $z = 3$ via $f_{\rm host}$ (and $M_{\rm min}$), so active quasars are always placed in ionized bubbles.

\emph{QSO1:} Halos are randomly populated with no preference to host mass. This method is inspired by  \citet{Hopk06}. In this picture, the distribution of quasar luminosities is due primarily to evolution in the light curve of individual black holes rather than differences in the black holes themselves: as quasar feedback clears out the host halo's gas, less of the light is obscured, rapidly increasing the flux reaching the IGM \citep{Hopk05, Hopk05a}. This scenario is consistent with quasars residing in halos of a characteristic size;  by choosing random halos, we preferentially populate the lowest halo mass, or $5\times 10^{11}$~\Msun, which is at the lower end of recent estimates (see, e.g., \citealt{Ross09, Trai12}). As shown above, a larger $M_{\rm min}$ would produce rarer and larger bubbles, but our conclusions would be minimally affected as detailed below.

\emph{QSO2:} In the second model, we assign the brightest quasars to the most massive halos, which is more consistent with the well-known $M$--$\sigma$ relation. Loosely, we are supposing that the luminosity of quasars increases with the halo mass, without fixing the exact form. This picture essentially assumes that the quasar's luminosity is proportional to its black hole's mass, and that the black hole mass increases monotonically with halo mass (see, e.g.,  \citealt{Wyit02}).  Any other factors -- such as the light curve of each quasar -- are subdominant in setting the relationship.

\emph{QSO3:} The last model assigns quasars to completely random positions within our box, irrespective of halo positions or ionized regions. This model does not include any clustering and serves mainly as an illustration.

\section{Photoionization rate} \label{sec:Gamma}

With the quasar configuration, ionization map, and density distribution, we can calculate the He~II photoionization rate $\Gamma$ at any point in our box, even during reionization. Although the mean rate $\langle \Gamma \rangle$ determines the average behavior of many observable quantities, we are particularly interested in the magnitude of fluctuations about this mean, which have important effects for scatter in observables and even for the evolution of the mean \citep{Davi12,Dixo13a}. In brief, we add up the contribution to $\Gamma$ from all sources within our box, taking into account the local density and the ionization state.

\subsection{Post-reionization} \label{sec:post}

First, we will examine the post-reionization regime, where the ionization maps are immaterial, and the variations in $\Gamma$ only depend on the distribution and intrinsic qualities of the active quasars. The specific intensity at a given frequency (in units of ergs/s/cm$^2$/Hz/sr) is
	\begin{equation} \label{eq:J} J_\nu~~=~~(1 + z )^2\sum_i\frac{L_{\nu,i}}{(4 \pi r^2_i)} e^{-r_i/\lambda_{\rm mfp}} , \end{equation} where $L_{\nu,i}$ is the $i^{\rm{th}}$ quasar's specific luminosity (converted using our assumed spectrum and including the dispersion in $\alpha$), $r_i$ is the comoving distance between this quasar and the point of interest, and $\lambda_{\rm mfp}$ is the comoving mean free path of the ionizing photons. Put simply, we add up the contribution of each quasar to the specific intensity at the point of interest, truncating at the box size. 

The mean free path is both uncertain and varies rapidly over the era of interest, with estimated values ranging from 15 to 150~Mpc at the ionization edge. We choose a fiducial value at this edge, $\lambda_0 = 60$~Mpc, to match more detailed calculations of $\lambda_{\rm mfp}$ that incorporate more sophisticated treatments of the \Lya forest \citep{Davi12}. However, it is worth noting that the mean free path varies quite rapidly with redshift over the $z \approx 2.5$--$3.5$ range and is subject to substantial uncertainties in the source population.  We allow a range of $\lambda_0 = 15, 35, 60, 80$~Mpc in this work to evaluate the impact.

At higher frequencies, we fix the mean free path as a power law in frequency,
	\begin{equation} \label{eq:R0} \lambda_{\rm mfp} = \lambda_0\left(\frac{\nu}{\nu_{\rm{HeII}}}\right)^{1.5}, \end{equation} 
where $\nu_{\rm{HeII}}$ is the frequency corresponding to the ionization edge of He~II. In a simple toy model for which the distribution of He~II absorbers $f(N_{\rm HeII}) \propto N_{\rm HeII}^{-\beta}$, we would have $\lambda_{\rm mfp} \propto \nu^{3(\beta - 1)}$ \citep{Fauc09}. Within the context of such a model, our slope corresponds to $\beta = 1.5$, which is close to the canonical value for H~1 absorbers \citep{Peti93, Kim02}.  Recent work shows that the H~1 \Lya forest column density distribution may be much more complex than such a toy model \citep{Proc10, OMea13, Rudi12a}, but detailed calculations of the ionizing background show that this frequency dependence is nevertheless accurate, at least relatively close to the ionization threshold \citep{Davi12}. We also explore a model in which the mean free path remains constant with frequency.

In detail, our assumption of a spatially constant mean free path is incorrect: because the ionization structure of the IGM absorbers is set by the (fluctuating) metagalactic radiation field, the amount of absorption -- and hence the mean free path -- fluctuates as well. We ignore this possibility here, but it has been approximately included in analytic models \citep{Davi12}.

\begin{figure*}
   \vspace{-5\baselineskip}
 {
    \includegraphics[trim = 0 0 5.5cm 10cm, clip,width=0.48\textwidth]{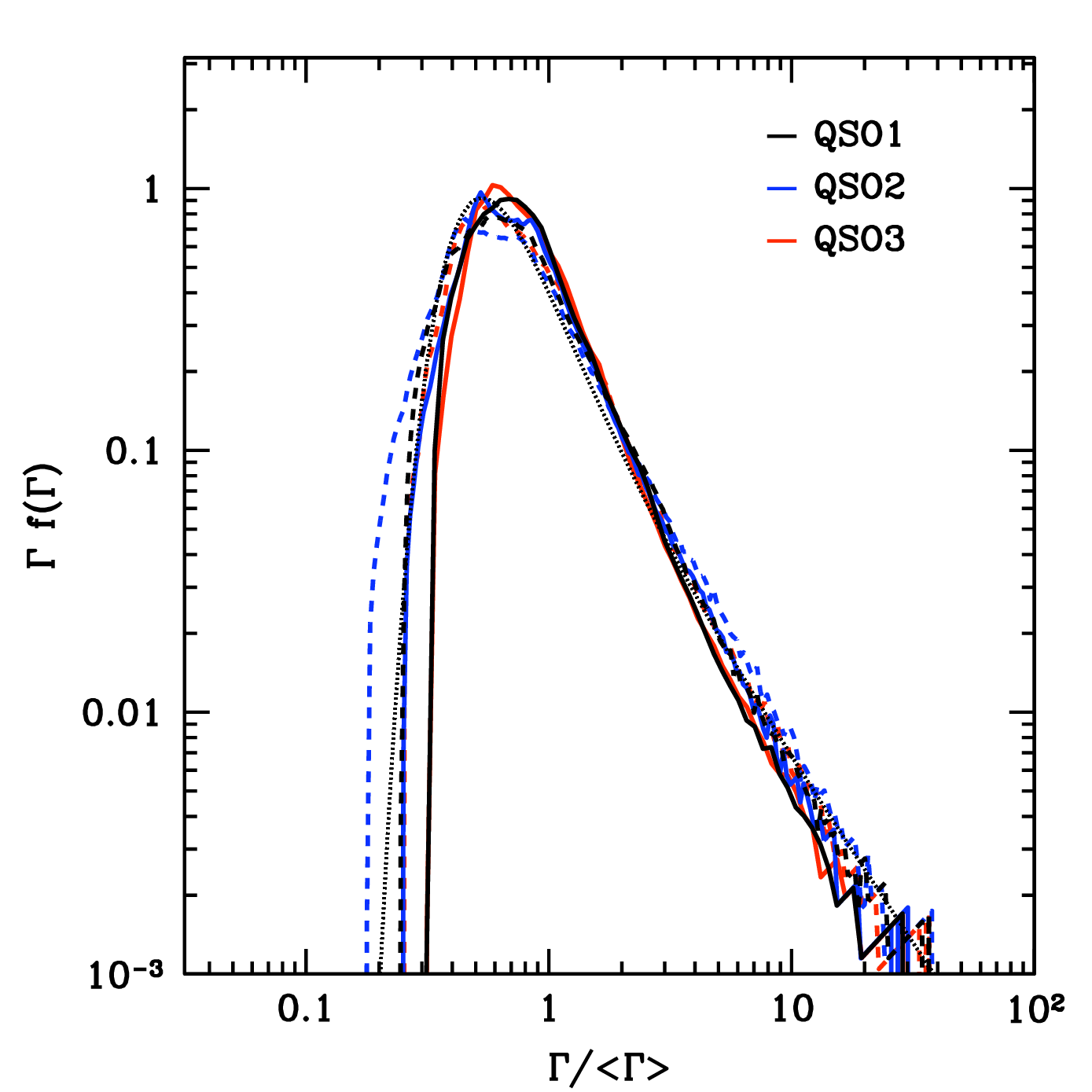}
   \includegraphics[trim = 0 0 5.5cm 10cm, clip,width=0.48\textwidth]{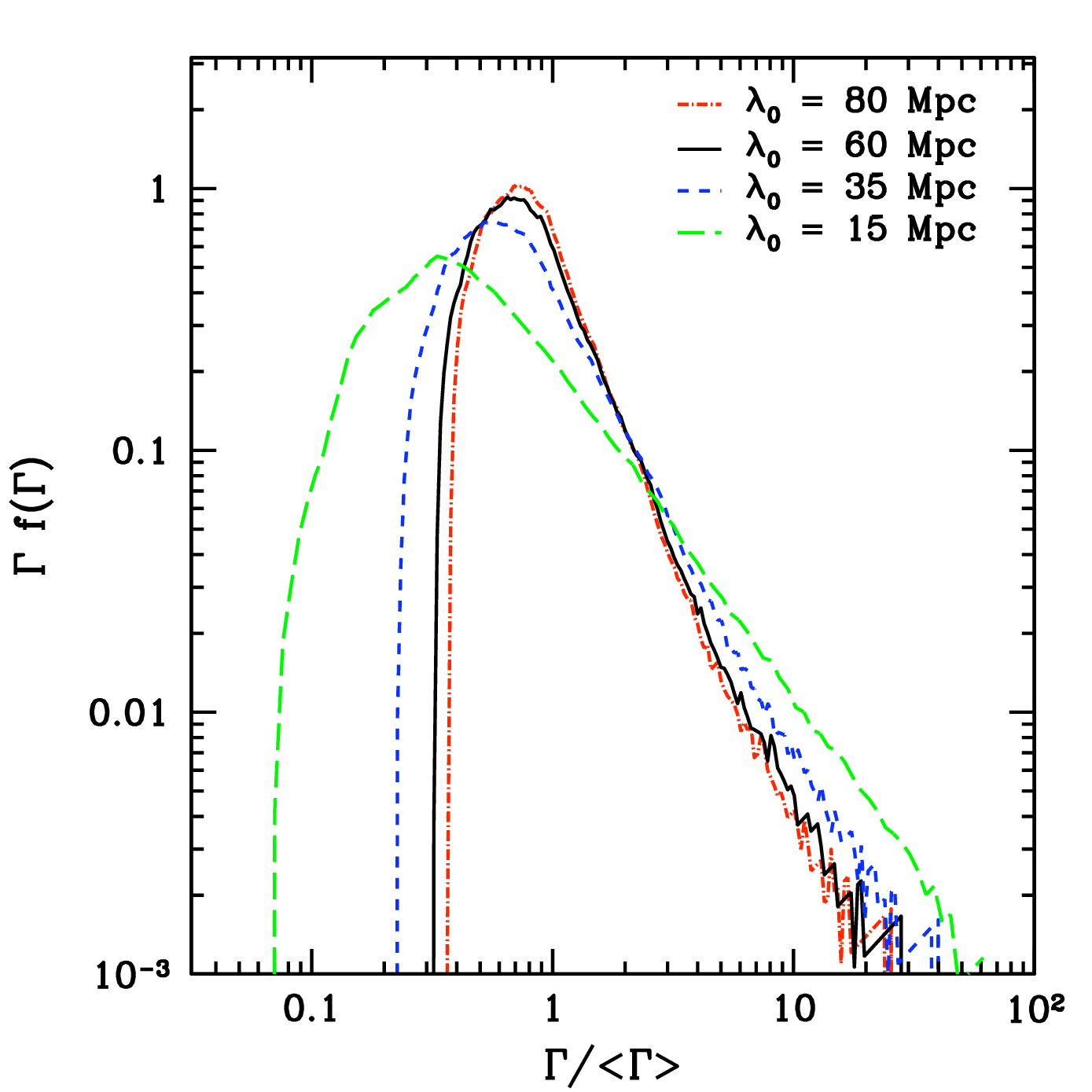} 
  }
  \caption{ Distribution of $\Gamma$ (scaled to the mean) in the post-ionization regime. \emph{Left panel:} Three models are shown: quasars placed in random halos (black), most luminous quasars situated in most massive halos (blue), and quasars placed randomly throughout the box (red). The dashed curves result from a fixed attenuation length, $\lambda_{\rm mfp} = 60$~Mpc. The post-reionization analytic result is represented by the dotted line. \emph{Right panel:} From widest to narrowest curves, $\lambda_0$ = 15, 35, 60, and 80~Mpc.}
   \label{fig:fG_QSO}
   \vspace{-1\baselineskip}
\end{figure*}

The photoionization rate follows simply from $J_\nu$:
	\begin{equation} \label{eq:Gint} \Gamma~~=~~4\pi \int_{\nu_{\rm{min}}}^{\infty} d\nu \frac{J_\nu}{h\nu}\sigma_{\rm{HeII}}(\nu), \end{equation} where $h$ is Planck's constant and $\sigma_{\rm{HeII}}(\nu) = \sigma_0(\nu/\nu_{\rm{HeII}})^{-3}$ ($\sigma_0 = 1.91\times10^{-18}~\rm{cm}^2$) is an approximation to the photoionization cross-section. In the post-reionization case, $\nu_{\rm{min}}$ is merely the photon frequency required to fully ionize helium. The probability distribution of $\Gamma$ is shown in the left panel of Fig.~\ref{fig:fG_QSO}, scaled to the average value for each scenario in a post-reionization universe. The difference between the three quasar models (\emph{QSO1} black, \emph{QSO2} blue, and \emph{QSO3} red) is fairly negligible in this regime. 
	
Since the results for \emph{QSO1} and \emph{QSO3} are nearly identical, randomly placed quasars approximates quasars placed in random halos, demonstrating that clustering of sources is unimportant. Predictably, the \emph{QSO2} result is wider, since clustering is more pronounced in this scenario. The largest dark matter halos tend to be found near other large dark matter halos, so the brightest quasars will tend to cluster. As noted in \citet{Furl08b}, these brightest quasars dominate the distribution. The $\langle 
\Gamma \rangle$ are identical to within a few percent for the three quasar models.

The left panel of Fig.~\ref{fig:fG_QSO} also compares frequency-dependent (solid) and frequency-independent (dashed) mean free paths. The differences between quasar models are similar in both cases, but the curves are generally wider for a frequency-independent value. Photons at the highest energies can travel further in the frequency-varying model providing a higher minimum ionizing background, although the difference is small in the post-reionization regime. Along similar lines, the right panel of Fig.~\ref{fig:fG_QSO} shows the impact of varying the mean free path normalization with $\lambda_0$ = 15, 35, 60, and 80~Mpc (widest to thinnest curves). As before, larger $\lambda_{\rm mfp}$ narrow the distribution. The larger $\lambda_0$ results are very similar, indicating that the quasars dominating the distribution are within $\sim\!60$~Mpc of each other; beyond that point, the intrinsic spread in luminosities dominates the distribution. 

The analytic model of \citet{Furl08b}, based on Poisson distributed sources \citep{Zuo92, Meik03} and represented by the dotted line, with the a fixed $\lambda_{\rm mfp}$ = 60~Mpc is remarkably close to the semi-numeric result. Since it does not include frequency dependence in $\lambda_{\rm mfp}$, the analytic result unsurprisingly matches the fixed $\lambda_{\rm mfp}$ model best. The congruence of the analytic and semi-numeric distributions further indicates that the rarity of quasars dominates the distribution of the photoionization rate, not clustering.

Even in this post-reionization regime, $\Gamma$ varies substantially about the mean, with very little dependence on the underlying details. The width of this distribution is essentially set by the QLF (constant for all models in this work) and the mean free path, where smaller values broaden the curve. In all cases, the high-$\Gamma$ tail is especially robust. This segment of the distribution corresponds to regions near bright sources, or proximity zones, so model details, like the mean free path or exact quasar placement, are inconsequential.

\subsection{During reionization} \label{sec:during}

Next, we approximate the evolution of $\Gamma$ during reionization (by varying \xHeIII). High-energy photons can propagate through the He~II regions present in this era, because the optical depth experienced by photons decreases with frequency.  Unlike previous analytic models, we can explicitly compute the ionizing background generated by these high-energy photons. In particular, with our density field and ionization map, we estimate the He~II column density $N_{\rm{HeII}}$ for a given photon ray and, thereby, estimate the absorption. 

Suppose we wish to calculate $\Gamma$ at a particular point ${\bf r}$.  Beginning there, we take a step of length $dx_p = \frac{1}{2} \Delta x_p$, where $\Delta x_p$ is the proper width of a pixel in our simulation. If the pixel contains He~II, the step's contribution to the intervening column density is $\Delta N_{\rm{HeII}} = \Delta_{\rho} \times n_{\rm{He}} \times dx_p$, where $\Delta_{\rho}$ is the pixel's overdensity and $n_{\rm{He}}$ is the (proper) number density of helium atoms. Otherwise, there is no contribution to $N_{\rm{HeII}}$. If the ionization state changes during a step, we use the \xHeII~present at the midpoint for the entire segment.
 
Given this column density, we can estimate the minimum photon frequency for light from a particular quasar $i$ to reach our point, $\nu_{\rm{min}}^i$, using the opacity condition $\sigma_{\rm{HeII}}(\nu_{\rm min}^i) N_{\rm{HeII}}^i = 1$, where $N_{\rm{HeII}}^i$ is the total column density between the point and the quasar.  Then, we demand
	\begin{equation} \label{eq:nu_min} \nu_{\rm{min}}^i~~=~~(\sigma_0 N_{\rm{HeII}}^i \nu_{\rm{HeII}}^3)^{1/3}. \end{equation} 
Of course, we require that $\nu_{\rm{min}} \geq \nu_{\rm{HeII}}$, irrespective of this equation. We then use this threshold frequency in eq.~(\ref{eq:Gint}) to calculate the total ionizing background, $\Gamma$.  In this way, we approximate the opacity due the He~II regions in a local, density-dependent manner. 

This component of the absorption typically dominates the opacity when \xHeIII~is small: at  \xHeIII~$\approx 0.50$, the ionized bubble size distribution, a reasonable proxy for the path length, peaks around 10~Mpc, which is much smaller than the mean free path we have assumed within the ionized gas. However, by \xHeIII~$\approx 0.80$, the peak is of the order of the mean free path.  Thus, photons typically reach $\tau \sim 1$ while traveling through nominally ionized regions during these late stages, indicating that dense clumps within these ionized regions become important before the completion of reionization.
	
\begin{figure*}
   \vspace{-5\baselineskip}
 {
    \includegraphics[trim = 0 0 5.5cm 10cm, clip,width=0.48\textwidth]{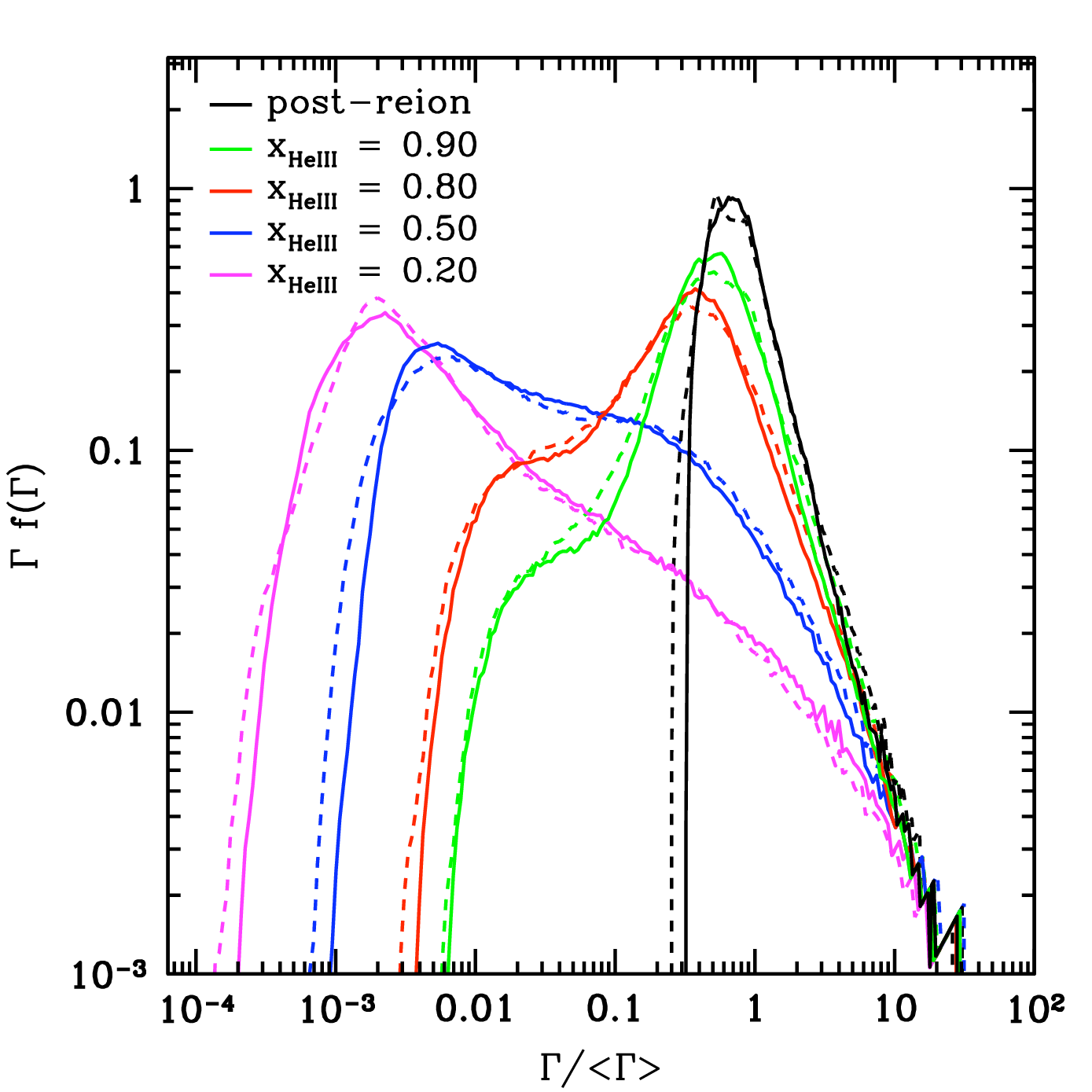}
   \includegraphics[trim = 0 0 5.5cm 10cm, clip,width=0.48\textwidth]{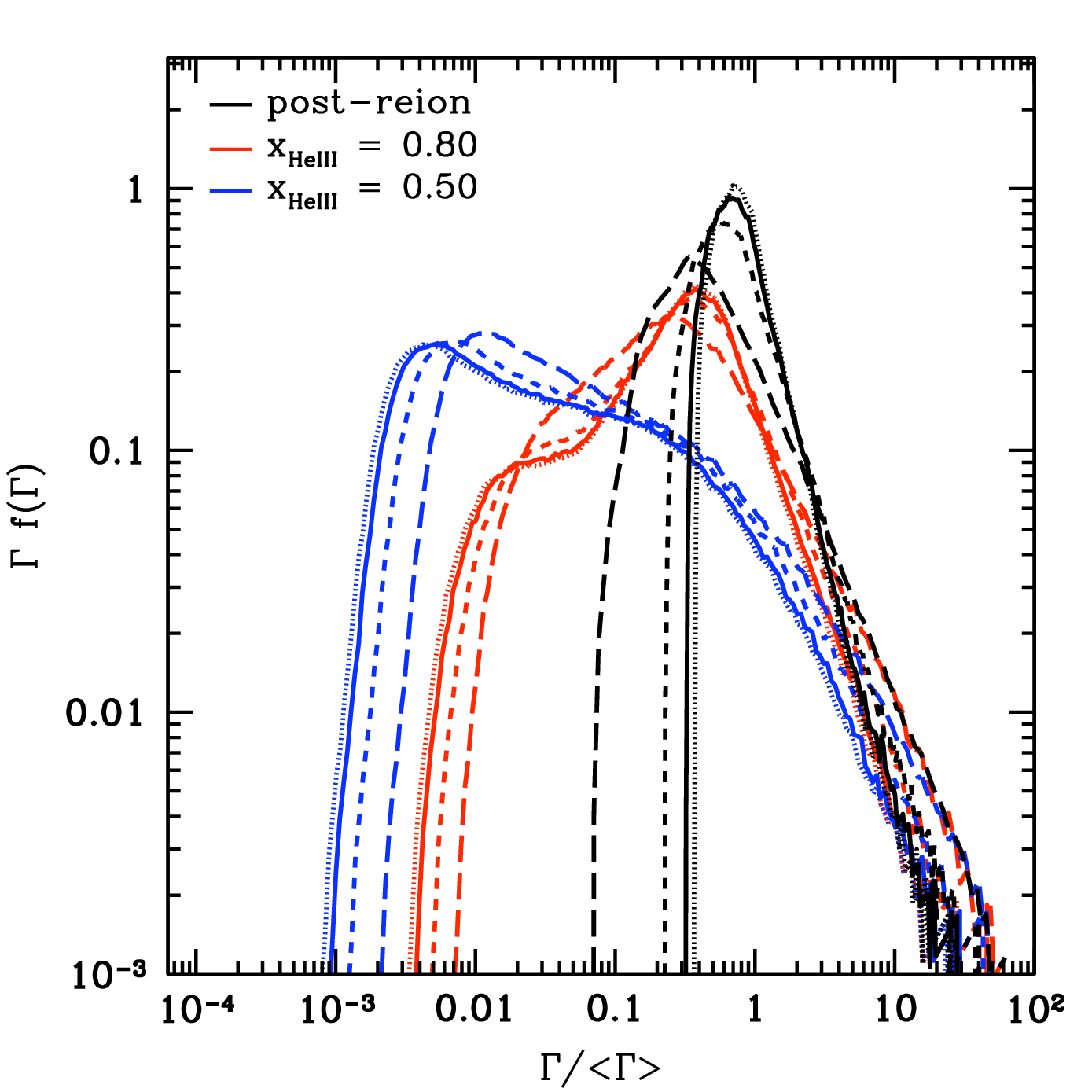} 
  }
   \caption{  Distribution of $\Gamma$ (scaled to the mean fiducial value post-reionization) for several \xHeIII~and varying $\lambda_0$. \emph{Left panel:} From lowest to highest at $\langle \Gamma \rangle$, \xHeIII~=  0.20, 0.50, 0.80, 0.90 and post-reionization are shown for the fiducial model (solid) and \emph{QSO2} variation (dashed). \emph{Right panel:} The fiducial model for \xHeIII~= 0.50, 0.80, and post-reionization is displayed with varied mean free path normalizations, $\lambda_0$ = 15, 35, 60, and 80~Mpc as long-dashed, short-dashed, solid, and dotted, respectively. Note that $\lambda_0$ = 60~Mpc (solid) is the fiducial value. 
     }
   \label{fig:fG_nf}
   \vspace{-1\baselineskip}
\end{figure*}

Fig.~\ref{fig:fG_nf} displays the distribution of $\Gamma$ for several \xHeIII. In the left panel, \emph{QSO1} (solid) and \emph{QSO2} (dashed) for the fiducial model are shown for \xHeIII~=  0.20, 0.50, 0.80, 0.90 and post-reionization (from widest to thinnest distributions, respectively). A bimodal distribution develops at small ionized fractions due to the low-level background from hard photons escaping into the He~II regions. Because He~II reionization causes substantial IGM heating, whose amplitude is determined by the shape of the local ionizing background, this bimodality may have important consequences for this heating process (\citealt{Furl08}; M09; \citealt{Bolt09a}). Even with \xHeIII~= 0.9, this hard background is still substantial. Interestingly, \citet{Mesi09} calculated the UV background using \textsc{DexM} during the H~1 reionization era.  At that time, the background does not exhibit this behavior. This difference is due to our inclusion of hard photons, which are more important for helium reionization (as the H~1 ionizing sources are expected to have had very soft spectra).

The choice of quasar model has only a minimal impact on the distributions, but it preferentially affects the high-end tail, which becomes increasingly important as \xHeIII~decreases. Put another way, the high-$\Gamma$ peak is shifted to the right for \emph{QSO2} for \xHeIII~$< 1$, while the low-$\Gamma$ peak is unaffected. Since \emph{QSO2} clusters more bright sources together, this shift indicates that proximity to quasars trumps the radiation from more distant sources. This behavior implies that detailed observations of He~II \Lya absorption inside of large ionized bubbles \emph{during} reionization may help determine how quasars are distributed inside of dark matter halos. 

The right panel of Fig.~\ref{fig:fG_nf} illustrates the effect of changing $\lambda_0$ on the distribution of $\Gamma$, where $\lambda_0$ = 15, 35, 60, and 80~Mpc as long-dashed, short-dashed, solid, and dotted, respectively, for \xHeIII~= 0.50, 0.80, and post-reionization. Note that the distribution is scaled to the post-reionization $\langle \Gamma \rangle$ for each $\lambda_0$. In the post-reionization regime, the general trend is that larger mean free paths translate into narrower distributions, as previously demonstrated. Similarly to post-reionization, the high-$\Gamma$ tail is steeper for larger $\lambda_0$ during reionization. In contrast, the low-$\Gamma$ peak does not shift to lower values for smaller $\lambda_0$, nor do the curves become significantly narrower. Since more photons can penetrate the He~II regions with larger mean free paths, the $\langle \Gamma \rangle$ of these regions (a proxy for the low-$\Gamma$ peak) is higher. To roughly summarize, the low peak and steepness of the high-$\Gamma$ tail are determined by how far photons can travel, and the high peak is shaped by the position of quasars (see \citet{Mesi08} for a similar conclusion for the hydrogen case).

\begin{figure*}
   \vspace{-5\baselineskip}
 {
    \includegraphics[trim = 0 0 5.5cm 10cm, clip,width=0.48\textwidth]{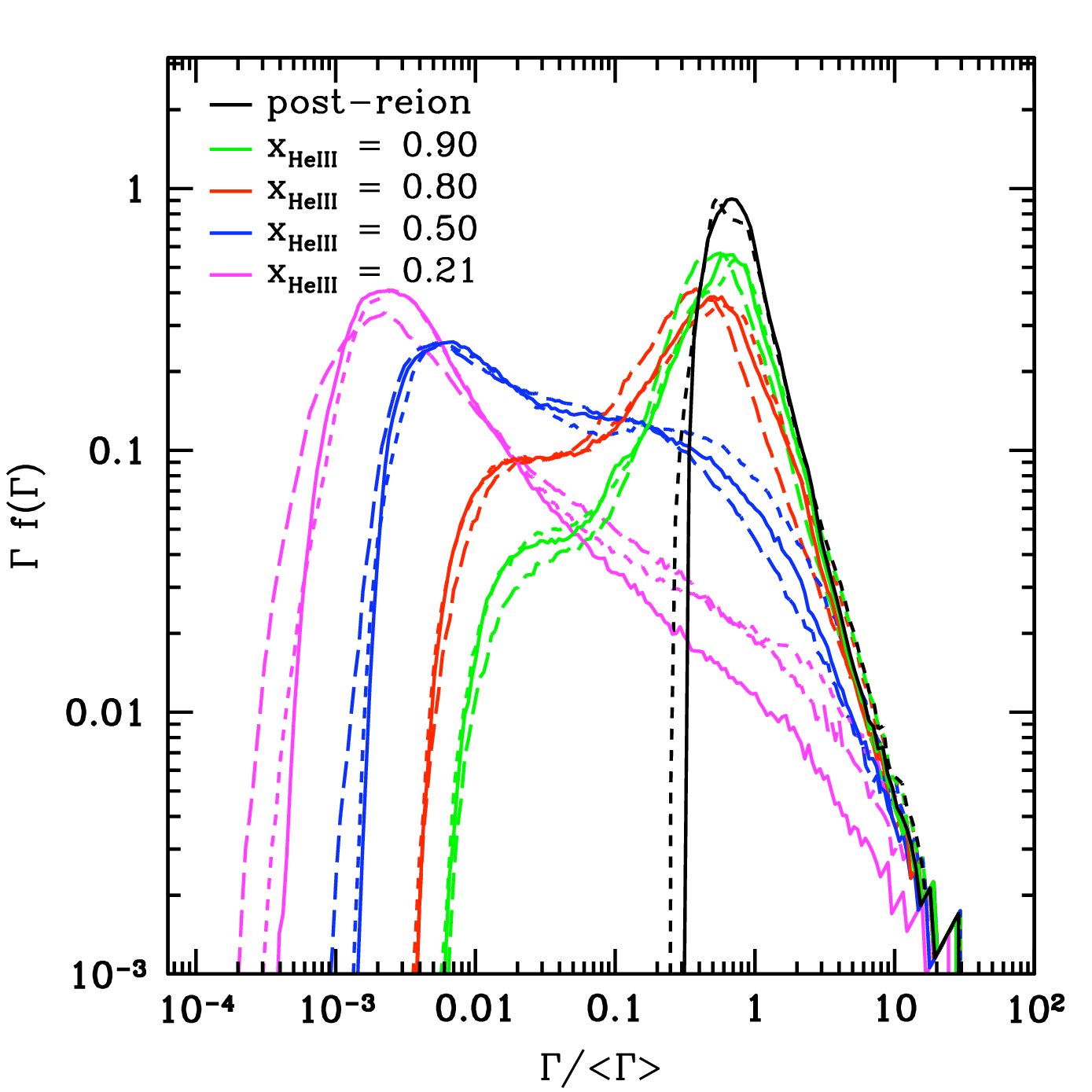}
   \includegraphics[trim = 0 0 5.5cm 10cm, clip,width=0.48\textwidth]{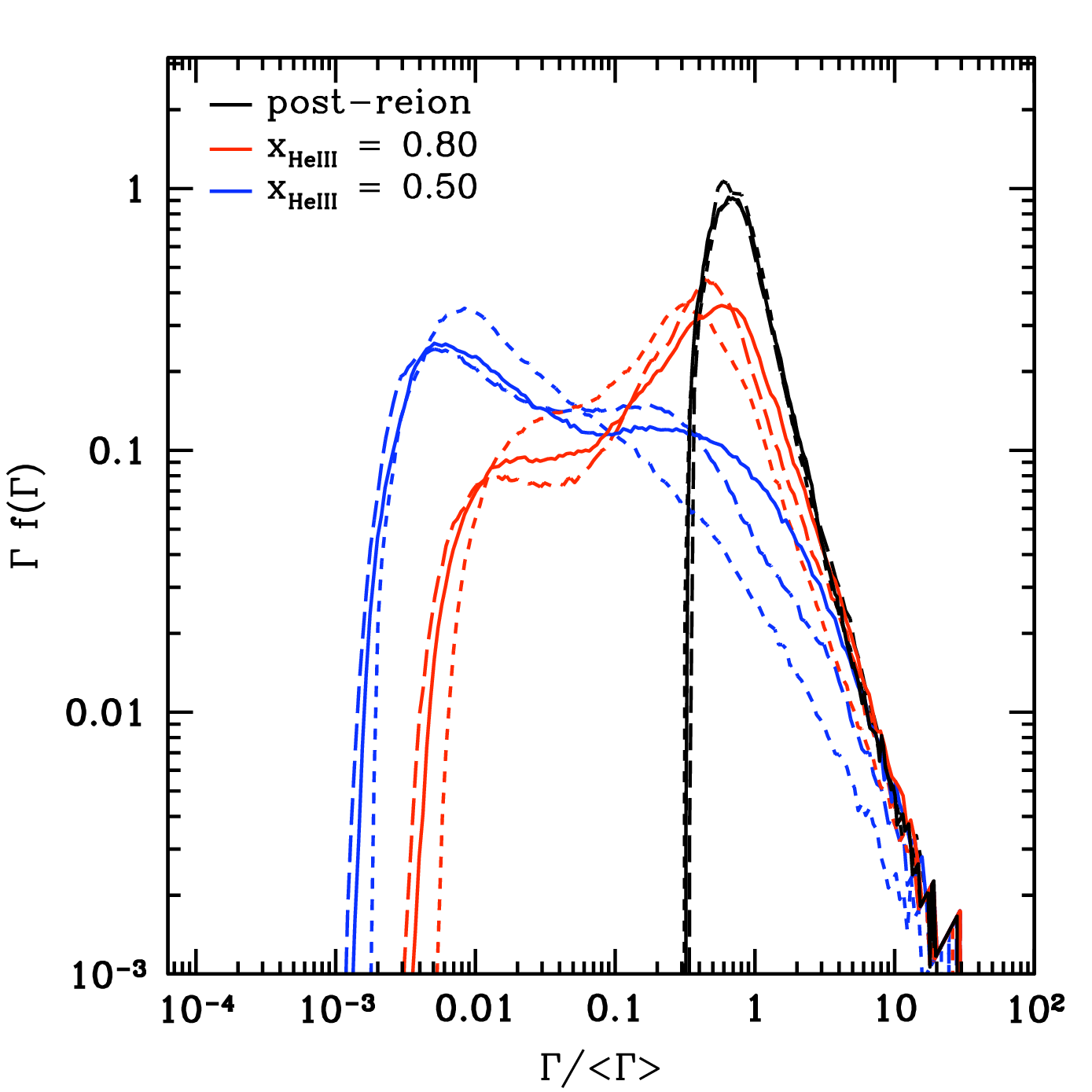} 
  }
  \caption{ Distribution of $\Gamma$ (scaled to the mean value post-reionization) for the abundant-source method. \emph{Left panel:} From lowest to highest at $\langle\Gamma\rangle$, \xHeIII~=  0.21, 0.50, 0.80, 0.90, and post-reionization, where \emph{QSO1} (solid) and \emph{QSO2} (short-dashed). For comparison, the long-dashed curves are the fiducial model with \emph{QSO1}. \emph{Right panel:} Here, the minimum halo mass is varied with \emph{QSO1}, and $M_{\rm min} = (0.5, 5, 10)\times 10^{11}~\Msun$ are the short-dashed, solid, and long-dashed curves, respectively. For clarity, only \xHeIII~=  0.50, 0.80, and post-reionization (from widest to narrowest) are shown.
   }
   \label{fig:fGa_nf}   \vspace{-1\baselineskip}
\end{figure*}

The left panel of Fig.~\ref{fig:fGa_nf} contrasts the abundant-source and fiducial methods. Here, \xHeIII~=  0.21, 0.50, 0.69, 0.80 and post-reionization (widest to thinnest curves, respectively) with \emph{QSO1} (solid) and \emph{QSO2} (short-dashed). In the abundant-source case, the difference between \emph{QSO1} and \emph{QSO2} is even less pronounced than in the fiducial framework. With smaller ionized bubbles, fewer pixels reside in an ionized region with an active quasar, except when overlap dominates (higher \xHeIII). Comparing the fiducial method with \emph{QSO1} (long-dashed curves), the abundant-source model are very similar for higher \xHeIII. For the lowest ionization fraction, the high-$\Gamma$ tail is substantially higher in the fiducial model, but the low-$\Gamma$ peak remains quite similar. The decrease at high intensities occurs for two reasons. First, the abundant-source method enforces smaller ionized bubbles, so the proximity zones are smaller and less space is filled by the near regions (outside the He~III region, of course, the opacity rapidly increases). Second, the abundant-source method has many more empty He~III regions -- that is, regions that have been ionized in the past but no longer host an active source.  Because more of the IGM is inside of isolated bubbles, there is less chance for a nearby quasar to illuminate such a region.  Overall, the difference between the two methods is surprisingly small, especially during the late stages of reionization.

The right panel of Fig.~\ref{fig:fGa_nf} shows the impact of the number of halos hosting sources (in the past) via the minimum halo mass $M_{\rm min} = (0.5, 5, 10)\times 10^{11}~\Msun$ as the short-dashed, solid, and long-dashed curves, respectively. Note that these are variations on the abundant-source model, where the representative ionization map slices are shown in Fig.~\ref{fig:variation}. The widest to thinnest curves are \xHeIII~= 0.5, 0.8, and post-reionization. In the post-reionization regime, the results are nearly identical, meaning the exact placement of active quasars is relatively unimportant. With smaller ionized bubbles (i.e., lower $M_{\rm min}$), the distributions are shifted toward lower $\Gamma$. Decreasing $M_{\rm min}$ reduces source clustering. Essentially, fewer regions are within an ionized bubble \emph{and} close to a quasar. 

Figs.~\ref{fig:fG_nf} and \ref{fig:fGa_nf} span a wide range of quasar model possibilities. Importantly, the density distribution (as noted in \S\ref{sec:density}), the exact ionization map, and even the basic quasar model do not greatly affect the photoionization rate distributions. Not only does this minimize the dependence on our fiducial choices, it emphasizes the robustness of our general conclusions and the relative importance of uncertain quantities relating to quasars.

\begin{figure}
   \vspace{-5\baselineskip}
 {
    \includegraphics[trim = 0 0 5.5cm 10cm, clip,width=0.48\textwidth]{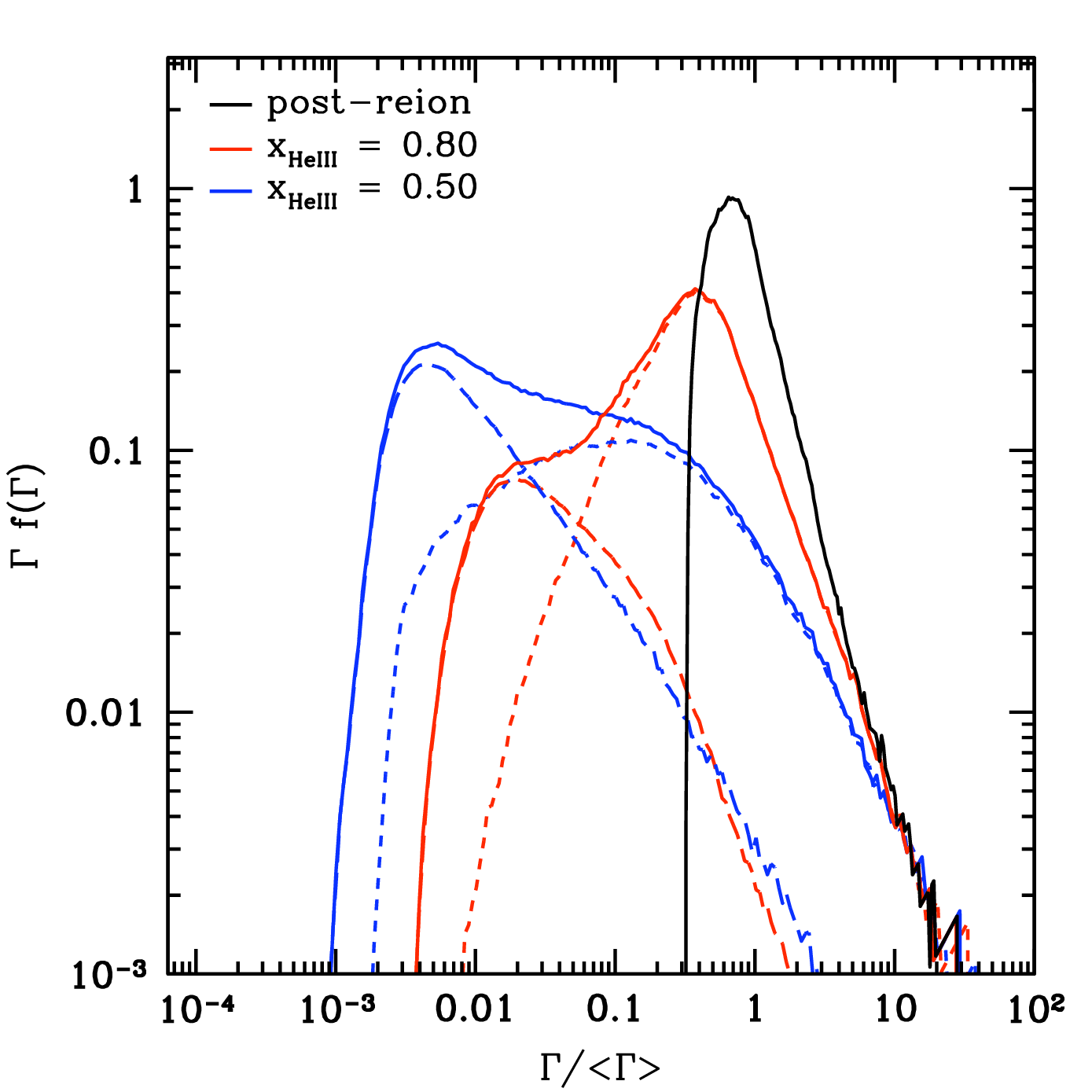}
  }
   \caption{ The distribution of $\Gamma$ within and without ionized regions. The solid, short-dashed, and long-dashed lines are the distributions of the entire box, He~III regions, and He~II regions, respectively. The widest to thinnest curves are \xHeIII~= 0.50, 0.80, and post-reionization. The high-end $\Gamma$ is determined almost exclusively by regions around quasars, which are necessarily inside ionized bubbles. The ionizing photons that escape the He~III dominate the low-end $\Gamma$, although ionized bubbles without an enclosed source also contribute. }
   \label{fig:fG_io}
   \vspace{-1\baselineskip}
\end{figure}

To tease out the cause of the bimodal nature of $\Gamma$, we contrast the distribution of $\Gamma$ inside fully ionized bubbles with that outside them in Fig.~\ref{fig:fG_io} (short and long-dashed curves, respectively). As expected, ionized areas around active quasars contain the highest $\Gamma$. The origin of the small-$\Gamma$ distribution is less straightforward. The ionizing radiation that penetrates the singly ionized helium comprises the majority of this low-end segment. This radiation is mainly due to high-frequency photons that can travel larger distances before absorption. Ionized bubbles that do not host \emph{active} quasars also make up a non-negligible fraction of this low-$\Gamma$ tail (and especially in the transition region from one peak to the other). This contribution is more relevant early in reionization (when overlap of the bubbles is less significant, meaning neighboring sources cannot illuminate such empty regions). We caution the reader that such empty regions are unlikely to be fully ionized, since the recombination time is relatively short \citep{Furl08c}. Therefore, we likely overestimate the ionizing background here.

\citet{Furl08b} explored the related $f(J)$ through a combination of analytic calculations and a Monte Carlo model for the number of sources within a distribution of bubble sizes. The results during reionization show large fluctuations but do not exhibit the bimodal behavior found in this work. The difference is due to our inclusion of local density effects and the improved treatment of hard photons via $\nu_{\rm min}$ and a frequency-dependent mean free path.  

As a simple exercise, let us use the peak of the low-$\Gamma$ distribution to estimate the path length of He~II through which photons typically travel.  In our model, that path length $\Delta R$ (here expressed in comoving units) determines $\nu_{\rm min}$ via eq.~(\ref{eq:nu_min}) and hence $\Gamma$ through eq.~(\ref{eq:Gint}).  If we assume that $J_\nu \propto \nu^{-\alpha}$ and $\alpha = 1.5$ for simplicity, we find that for a given ionization rate $\Gamma$
\begin{equation}
\Delta R \sim 0.34 \left( {\Gamma_{\nu_{\rm min}=\nu_{\rm HeII}} \over \Gamma} \right)^{2/3} \left( {4 \over 1+z} \right)^2 \ {\rm Mpc},
\end{equation}
where $\Gamma_{\nu_{\rm min}=\nu_{\rm HeII}}$ is the corresponding ionization rate in a fully ionized medium transparent to all photons above the ionizing edge.  Using the peak of the high-$\Gamma$ distribution as a proxy for this value, our curves have $\Delta R \sim 10$~Mpc throughout the bulk of reionization, indicating that most He~II regions are quite large.

It is interesting to note that the low-level ionizing background present primarily in the He~II regions is nearly capable of fully reionizing helium, but on a very large timescale. To illustrate this, enforcing ionization equilibrium (assuming a clumping factor near 1) means the He~III fraction is
\begin{equation} x_{\rm HeIII}  ~~\approx~~ 1 - \frac{\alpha_{\rm B}X\Omega_B\rho_{crit}}{\Gamma m_h}, \end{equation}
where $X$ is the hydrogen mass fraction, $\alpha_{\rm B}$ is the recombination coefficient, $\rho_{crit}$ is the critical density, and $m_h$ is the mass of hydrogen. Taking $\Gamma$ as the low peak for $50$ per cent ionized, we find \xHeIII~$\sim$~0.98. The ionization timescale at this level -- about 1~Gyr -- is likely longer than the entire duration of helium reionization, meaning that reionization is driven primarily by ionized bubbles.

\section{Comparison with H~1 Reionization} \label{sec:hydrogen}

It is useful to consider the similarities and differences between our work here and analogous studies of H~1 reionization, especially given that we are adapting a hydrogen reionization code.  In this section, we identify the major points of divergence between these two epochs.

Our semi-numeric approach was originally designed for calculations of H~1 reionization, to which it is undoubtedly better suited.  Most importantly, the two-phase approximation (fully ionized or fully neutral) is very accurate for H~1 reionization, because the soft stellar sources most likely responsible for H~1 reionization have very short mean free paths ($\la$ a few kpc).  The ``photon-counting" arguments inherent to the method are therefore almost exactly accurate for that case, whereas the transition region between He~II and He~III fills a non-negligible fraction of the volume.  Our ionization maps must therefore be taken as approximate, although they are still useful for many purposes (such as calculations of the He~II \Lya optical depth, which is very large even in the transition region).

Another key difference between the ionization fields in H~1 and He~II reionization is the source abundance. We typically assume that any dark matter halo able to form stars will also produce ionizing photons during the earlier era, making those sources many times more common than luminous quasar hosts at $z \sim 3$.  During H~1 reionization, ionized bubbles typically have thousands of sources even early in the process \citep{Furl04}, but we have found that, for He~II reionization, even a single source can generate a large bubble.  The ``morphology" of reionization (i.e., the distribution of ionized bubble sizes)  is often taken as a key observable of the H~1 process, because the huge number of sources closely connects it to the underlying dark matter field.  In the He~II case, however, the small numbers of sources make that connection mostly uninteresting: Poisson fluctuations instead play a key role \citep{Furl08}, and the distribution of bubbles relative to halos is much more likely to tell us about the details of the black hole-host relation than it is the fundamentals of the reionization process.  Quantitatively, there is little evolution in the shape of the power spectrum of the ionized fraction throughout helium reionization.

Additionally, because there are so many sources inside each H~II region, the ionizing background remains large at all times, preventing these bubbles from recombining substantially.  When only one or two (short-lived) sources occupy each bubble, however, recombinations can substantially increase the He~II fraction inside even regions that are nominally fully ionized \citep{Furl08c}.  Fortunately, Fig.~\ref{fig:fG_io} shows that most such regions will be illuminated, at least late in the process.  But this will, for example, increase the opacity of some He~III bubbles in the \Lya forest (and to ionizing photons).  We have crudely modeled this effect through our source prescriptions (one can regard turning some halos ``off" as instead allowing their ionized bubbles to recombine fully).

Our approach to describing the ionizing sources themselves is also rather different.  The lack of data on high-redshift galaxies so far implies that naive prescriptions (with a constant ionizing efficiency per halo, or at most a systematic variation with halo mass) are more than adequate.  On the other hand, the tight observational constraints on the QLF at $z \sim 3$ require that it be directly incorporated into the calculation.  This inclusion is easy to do for the instantaneous ionization rate but much harder for the cumulative number of ionizing photons, because the relationship between the visible sources and their dark matter halo hosts and the light curves of individual quasars remains uncertain.  We have shown that the details of this association do not strongly affect many properties of reionization (such as the distribution of $\Gamma$), but they will show up in other observables.  For example, one could imagine searching for massive halos inside of large He~III regions in order to identify the descendants of bright quasar hosts.

A final key difference is the source spectrum, which is unimportant for stellar sources but crucial for the quasars responsible for He~II reionization.  Although we have ignored the resulting partially ionized transition zones, we have included these spectra explicitly in evaluating the ionizing background.  We are thus able to track the growth of a hard background in He~II regions, an effect likely to be much less important during H~II reionization \citep{Mesi09}.  

\section{Discussion} \label{sec:disc}

We have introduced a new and efficient semi-numeric method for efficiently generating dark matter halo distributions and ionization maps relevant to the full reionization of helium. Specifically, we adapted an existing hydrogen reionization code (\textsc{DexM}) to identify halos at lower redshift and changed the method for determining the ionization state to suit quasars as sources. The speed of our algorithm allows exploration of a wide range of possible quasar models not only for active quasars, but in the realization of the ionization map as well. 

Our results broadly match previous simulations of related quantities, at least statistically speaking. We generated accurate halo mass functions as compared to $N$-body simulations. This congruence is important for determining the effect of quasar clustering. We also reproduced the He~III morphology, especially at high \xHeIII, as demonstrated by the ionized power spectra, which compare favorably with M09 (modulo our different source prescriptions). We demonstrated that this method can easily produce simulation volumes hundreds of Mpc across to accurately capture the large-scale features of helium reionization. We also constructed a density field that matches (by design) the smoothed particle hydrodynamics simulations of BB09. A realistic density field is crucial to capturing at least some of the small-scale fluctuations in the observable quantities, which is lacking in semi-analytic models.

We have presented the first systematic study of fluctuations in the ionizing background during reionization. Our $f(\Gamma)$ exhibit significant variation even post-reionization, in excellent agreement with analytic calculations \citep{Furl08b}. Furthermore, we calculate the more relevant $f(\Gamma)$ as opposed to $f(J)$ in the analytic case. We demonstrated that complications like a frequency-dependent mean free path and quasar clustering do not significantly affect the result in this regime. 

We also find that a broader and bimodal distribution develops during reionization. This distribution differs from that during the hydrogen reionization era \citep{Mesi08} and from analytic predictions for the He~III era \citep{Furl08b}. These fluctuations have important consequences, which we can quantify, for many observables, including the quasar spectra metal lines and IGM heating. In an upcoming paper, we will study the impact of these fluctuations on the He~II \Lya forest. The most important conclusion is that not every part of the IGM experienced the same reionization history or receives the same background radiation.

We included a frequency-varying mean free path, another crucial improvement on semi-analytic models. This frequency-dependent spectrum gives rise to the bimodal nature of $f(\Gamma)$ we find during reionization. Specifically, hard photons can travel farther and escape into the He~II regions, creating a low-level background there. This frequency-dependence is important for studies of the He~II \Lya forest, metal lines, and IGM heating.

There are some important caveats to our methods: 

(\emph{i}) We assign the quasars a simple -- though frequency-dependent -- mean free path with an uncertain normalization. This assumption ignores detailed radiative transfer effects and is important both during and after reionization.  In particular, we \emph{prescribe} the normalization of this mean free path, rather than allowing it to vary with the local radiation field \citep{Davi12}.

(\emph{ii}) We assume a two-phase medium. In principle, our ionized bubble edges should not be abrupt, especially since helium recombines so quickly, but instead transition smoothly from fully ionized to singly ionized. This effect is less important during the late stages of reionization as the bubbles increasingly overlap (so that the ratio of their volume to their surface area decreases).

(\emph{iii}) We do not fully account for recombinations. As mentioned above, we do not explicitly include any He~II inside ionized bubbles, though likely some He~III would have recombined. Not only would this affect our effective mean free path (as determined by $\nu_{\rm min}$), we do not compensate for ionizing photons being ``wasted" on reionizing this gas when generating our ionization maps.

(\emph{iv}) Our simulations assume that quasars emit radiation isotropically. Many theoretical models pertaining to quasar behavior predict otherwise. Although this effect is important during the early stages of reionization, once each ionized bubble contains (or has contained) multiple, randomly oriented quasars, beaming is no longer important.

Our conclusions are mostly insensitive to our underlying assumptions for active quasars and ionizing sources. For our wide range of explored models, the resultant $f(\Gamma)$ varied minimally, except for $\lambda_0$. In particular, quasar clustering is a secondary effect, and a few rare, bright quasars dominate the spectrum. Although this makes our predictions robust, distinguishing between our models using observational data dependent on the ionizing background would be difficult if not impossible.

With the ionization maps, density field, and active quasar models, we are uniquely suited to study fluctuations in the He~II \Lya forest. In particular, we maintain spatial information that is important for considering correlations along a line of sight. We explore these fluctuations in an upcoming paper \citep{Dixo13a}, where we show that the large variations in transmission indicate ongoing reionization at $z \gtrsim 2.8$. Similarly, these methods could apply to the He~II Lyman-$\beta$ forest, as measured by \citet{Syph11a}.

Further applications that we intend to explore are the transverse proximity effect, as measured in the He~II \Lya forest \citep{Jako03, Wors06, Wors07}, whereby a quasar near another line of sight enhances the ionizing background. The size and nature of the apparent influence of the \Lya spectra may give important clues about the intrinsic properties of quasars (e.g., \citealt{Furl11}). Also, the spatial inhomogeneities of the UV radiation background should  affect the ionization balance of the metals in the IGM. As an example, there may be a break in the ratio of C~IV to Si~IV at $z \sim 3$ indicating helium reionization (\citealt{Song98, Song05}, though see \citealt{Agui04}). The fluctuations we find in our He-ionizing background could have important consequences for these measurements. Combining past and upcoming observations with our theoretical tools will shed light not only on the timing and duration the helium reionization epoch, but also the nature of the IGM and the properties of quasars.

\section{Acknowledgments} \label{sec:thanks}

We thank F.~Davies for helpful discussions and sharing preliminary results. This research was partially supported by the David and Lucile Packard Foundation, the Alfred P. Sloan Foundation, and NSF grant AST-0607470. KLD gratefully acknowledges support by the Science and Technology Facilities Council grant number ST/I000976/1.

\footnotesize{
\bibliographystyle{mn2e}
\bibliography{../helium_reionization}
}

\end{document}